\documentclass[12pt,a4paper]{article}
\usepackage[cp866]{inputenc}
\usepackage{amsmath}
\usepackage{axodraw}
\usepackage[dvips]{epsfig}
\usepackage{graphicx}
\usepackage{epsfig}
\newcommand{\be}{\begin{equation}}
\newcommand{\ee}{\end{equation}}
\newcommand{\ben}{\begin{equation*}}
\newcommand{\een}{\end{equation*}}
\newcommand{\bea}{\begin{eqnarray}}
\newcommand{\eea}{\end{eqnarray}}
\newcommand{\ar}{\begin{array}}
\newcommand{\arn}{\end{array}}

\newcommand{\vk}{\vec{k}}
\newcommand{\vks}{\vec{k}^{\;2}}
\newcommand{\q}{\vec{q}}
\newcommand{\qs}{\vec{q}^{\;2}}
\newcommand{\qp}{\vec{q}^{\;\prime}}

\newcommand{\p}{\vec{p}}

\newcommand{\vl}{\vec{l}}
\newcommand{\vls}{\vec{l}^{\;2}}

\newcommand{\x}{\vec{r}}
\newcommand{\xs}{\vec{r}^{\;2}}
\newcommand{\xp}{\vec{r}^{\;\prime}}
\newcommand{\xps}{\vec{r}^{\;\prime\; 2}}

\def\pnot{\mbox{${\not{\hbox{\kern-3.0pt$p$}}}$}}
\def\qnot{\mbox{${\not{\hbox{\kern-2.0pt$q$}}}$}}
\def\enot{\mbox{${\not{\hbox{\kern-2.0pt$e$}}}$}}
\def\knot{\mbox{${\not{\hbox{\kern-2.0pt$k$}}}$}}

\def\fun#1#2{\lower3.6pt\vbox{\baselineskip0pt\lineskip.9pt\ialign
{$\mathsurround=0pt#1\hfil##\hfil$\crcr#2\crcr\sim\crcr}}}

\begin{document}
\numberwithin{equation}{section}     
\sloppy                              
\renewcommand{\baselinestretch}{1.0} 

\begin{titlepage}

\hskip 11cm \vbox{ \hbox{Budker INP 2015-9}  }
\vskip 3cm
\begin{center}
{\bf  Discontinuites of BFKL amplitudes and the BDS ansatz}
\end{center}

\centerline{V.S.~Fadin
$^{a\,\dag}$, R.~Fiore$^{b\,\ddag}$}

\vskip .6cm

\centerline{\sl $^{a}$
Budker Institute of Nuclear Physics of SD RAS, 630090 Novosibirsk
Russia}
\centerline{\sl and Novosibirsk State University, 630090 Novosibirsk, Russia}
\centerline{\sl $^{b}$ Dipartimento di Fisica, Universit\`a della Calabria,}
\centerline{\sl and Istituto Nazionale di Fisica Nucleare, Gruppo collegato di Cosenza,}
\centerline{\sl Arcavacata di Rende, I-87036 Cosenza, Italy}

\vskip 2cm

\begin{abstract}
We perform an  examination of discontinuities of  multiple production amplitudes, which are required for further development of the BFKL  approach.
It turns out that the  discontinuities of $2\rightarrow 2+n$ amplitudes  obtained in the BFKL approach contradict to  the BDS ansatz for amplitudes  with maximal helicity violation in ${\cal N}=4$ supersymmetric Yang-Mills theory with large number of colours starting with $n =2$.  Explicit expressions  for the discontinuities  of the $2\rightarrow 3$ and  $2\rightarrow 4$ amplitudes  in  the invariant mass of pairs of  produced gluons are  obtained  in the planar N=4 SYM in the next-to-leading logarithmic approximation. These expressions   can be used for checking the conjectured duality between the light-like Wilson loops and the MHV amplitudes.
\end{abstract}


\vfill \hrule \vskip.3cm \noindent $^{\ast}${\it Work supported 
in part by the Ministry of Education and Science of Russian Federation,
in part by  RFBR,  grant 13-02-01023, and in part by Ministero Italiano dell'Istruzione,
dell'Universit\`a e della Ricerca.} \vfill $
\begin{array}{ll} ^{\dag}\mbox{{\it e-mail address:}} &
\mbox{fadin@inp.nsk.su}\\
^{\ddag}\mbox{{\it e-mail address:}} &
\mbox{roberto.fiore@cs.infn.it}\\
\end{array}
$
\end{titlepage}

\vfill \eject

\section{Introduction}

The BFKL (Balitsky-Fadin-Kuraev-Lipatov) approach  \cite{Fadin:1975cb, Kuraev:1976ge, Kuraev:1977fs, Balitsky:1978ic} is based on the multi-Regge form of scattering amplitudes with gluon quantum numbers in all cross-channels. For the amplitude  ${\cal A}_{2\rightarrow n+2}$ of  the  process $A+B\rightarrow A'+G_1+\ldots+G_n+B'$  of production of $n$ gluons with momenta $k_1, k_2, \ldots k_n $ in the multi-Regge kinematics (MRK)
this form   can be written as
\be
\Re {\cal A}_{2\rightarrow n+2}=2s\Gamma^{{R}_1}_{
A' A} \left( \prod_{i=1}^{n}
 \frac{1}{t_{i}}\Big(\frac{s_i}{|\vk_{i-1}||\vk_{i}|}\Big)^{\omega(t_i)}
\gamma^{G_i}_{R_i{R}_{i+1}}\right)\frac{1}{t_{n+1}}\Big(\frac{s_{n+1}}{|k_{n}||\q_{n+1}|}\Big)^{\omega(t_{n+1})}
\Gamma^{{R}_{n+1}}
_{B' B},
\label{A 2-2+n}
\ee
where $\omega(t)$ is called  gluon trajectory (in fact, the trajectory is $1+\omega(t)$),  ${\Gamma}^{{R}}_{ A'A}$ and $\Gamma^{{R}} _{B' B}$ are the particle-particle-Reggeon (PPR) vertices, or the  scattering vertices,  and  $\gamma ^{G_{i}}_{{R}_i {R}_{i+1}}$ are the Reggeon-Reggeon-gluon (RRG) vertices, or the production vertices. Moreover
\[
s=(p_A+p_B)^2, \; s_i=(k_{i-1}+k_i)^2, \;\;i=1, \cdot\cdot\cdot n+1, \;   \; k_{0}\equiv P_{A'}, \;\;k_{n+1}\equiv P_{B'},
\]
\be
q_1=p_A-p_A',\;   q_{j+1} = q_{j}-k_j, \; j=1, \cdot\cdot\cdot n, \; \; q_{n+1} =p_{B'} -p_B~,  \label{notation}
\ee
the vector sign means transverse to the   $p_A, p_B$ plane  components. In the MRK
\be
s \gg s_i \gg |t_i|\simeq \qs_i,~~~~ \;\; s\simeq \frac{ \prod_{i=1}^{n+1}s_i} {\prod_{i=1}^{n}\vks_i}~.\label{MRK}
\ee
The Reggeon vertices and the gluon trajectory are known  in the next-to-leading order (NLO), that means the one-loop approximation for the vertices and the two-loop approximation for the trajectory, in   $SYM$ as well as in QCD. It is just the accuracy which is required for the derivation of the BFKL equation in the next-to-leading logarithmic approximations (NLLA), taking into account  all radiative corrections of the type $\alpha_s\left(\alpha_s\ln s\right)$. To be precise, note   that in this approximation one has to consider not only the  amplitudes  \eqref{A 2-2+n},   but also amplitudes obtained from them by  replacement of one of final particles by  a couple of particles with fixed (of order of transverse momenta) invariant  mass.

The sign $\Re $ in the Eq.~\eqref{A 2-2+n} means the real part. It is important that this simple factorized  form is valid only for the real part of the amplitudes. Fortunately, the imaginary parts  are  not essential  for  the derivation of the BFKL equation in the NLLA, because they are  suppressed by one power of $\ln s_i$ in comparison with the real ones, and with the NLLA accuracy do not  contribute in the unitarity relations.  But understanding
of properties of the imaginary parts which are associated with the discontinuities in  the variables $s_{ij}=(k_i+k_j)^2$ is very important. First, it is necessary for the justification of the BFKL approach, that means a proof of the multi-Regge form of multiple production amplitudes.  Second, account of  the imaginary parts is  indispensable in further development of the BFKL approach. As it was pointed above, they are  not essential for derivation of the BFKL equation in the NLLA, but they must be taken into account in the NNLLA.

The   idea  of the multi-Regge form   appeared in  Refs.~\cite{Fadin:1975cb, Lipatov:1976zz}   from results of fixed order calculations.  Later  it was proved in the leading logarithmic approximation (LLA) \cite{Balitskii:1979} with use of  the  $s$-channel unitarity. The proof of the  multi-Regge form  in the  NLLA  is based also on the  $s$-channel unitarity \cite{Fadin:2006bj}.

 Here it is necessary to recall that as compared with ordinary particles, Reggeons
in the Regge-Gribov theory of complex angular momenta  possess an additional quantum  number, called signature. At large $s_i$ the signature means parity with respect to the substitution $s_i\rightarrow -s_i$. The signature of the Reggeized gluon is negative, and the real part of the amplitude presented in Eq.~\eqref{A 2-2+n} coincides  with the real part of the amplitude ${\cal A}^{\{-\}}_{2\rightarrow 2+n}$ with the Reggeized gluons (and, consequently, with the negative signatures) in all $t_i$ channels.  Amplitudes with the positive signature in the $s_i$-channel are  suppressed because of the cancellation of leading powers of $\log s_i$, so that with the NLLA accuracy $\Re {\cal A}_{2\rightarrow 2+n} =\Re{\cal A}^{\{-\}}_{2\rightarrow 2+n}$.

Compatibility of unitarity with the multi-Regge form leads to the bootstrap relations \cite{Fadin:2002et} connecting discontinuities of the amplitudes
with products of their real parts and gluon trajectories:
\be
\frac{1}{-\pi i}\left(\sum_{l=j+1}^{n+1}\mathrm{disc}_{s_{j,l}}
-\sum_{l=0}^{j-1}\mathrm{disc}_{s_{l,j}}\right){\cal A}^{\{-\}}_{2\rightarrow n+2} \, =\,
\left(\omega(t_{j+1})-\omega(t_{j})\right)\Re {\cal A}_{2\rightarrow n+2}~.
\label{bootstrap relations}
\ee
Here $\Re {\cal A}_{2\rightarrow n+2}$ is the multi-Regge form \eqref{A 2-2+n} and the $s_{ij}$-channel discontinuities  must be calculated using  this form
into the unitarity conditions. Note that for  multi-particle amplitudes the discontinuities are not pure imaginary, since a discontinuity
in one of the channels can have, in turn, a discontinuity in another channel. But
these double discontinuities are sub-sub-leading, so that they are  neglected in Eq.~\eqref{bootstrap relations}   and  in the following.

It turns out \cite{Fadin:2006bj} that  the fulfilment of an infinite set of the relations \eqref{bootstrap relations} guarantees the multi--Regge form of scattering  amplitudes and that all bootstrap relations  are fulfilled if several conditions imposed on the Reggeon vertices and the trajectory (bootstrap conditions) hold true.   The most complicated condition, which includes the impact factors for Reggeon-gluon transition, was
proved  recently,  both in QCD \cite{Kozlov:2011zza}-\cite{Kozlov:2012zza} and in its supersymmetric generalisations \cite{Kozlov:2013zza}.

The proof that  the fulfilment of the  bootstrap  relations \eqref{bootstrap relations} is ensured by the bootstrap  conditions is based on the form of the discontinuities derived from the unitarity in Ref.~\cite{Fadin:2006bj}. Besides of the Reggeon vertices and the trajectory entering in Eq.~\eqref{A 2-2+n}, the discontinuities contain as building blocks  the impact factors for particle-particle and Reggeon-particle transitions,  the kernel of the BFKL equation and the four-Reggeon gluon production vertex. In fact, the bootstrap conditions are conditions  on these building blocks. But since the impact factors for particle-particle and Reggeon-particle transitions,  the kernel of the BFKL equation and the four-Reggeon gluon production vertex are expressed in terms of the Reggeon vertices and the trajectory, one can say that the bootstrap conditions are imposed on the Reggeon vertices and the trajectory.

The expressions for  discontinuities obtained in  Ref.~\cite{Fadin:2006bj}  are rather formal,  since the  impact factors,  the kernel and the four-Reggeon gluon production vertex are not given explicitly.  In this paper we obtain explicit expressions for the discontinuities  of multiple production amplitudes in $N=4$ SYM with large number of colours (in the planar approximation). Consideration of the discontinuities in this theory is also interesting  for two reasons. First, it provides a  simple demonstration of imperfection  of the BDS (Bern-Dixon-Smirnov) ansatz  \cite{Anastasiou:2003kj, Bern:2005iz}  $M_{BDS}$ for multi-particle   amplitudes with maximal helicity violation (MHV amplitudes).
 Second,  the discontinuities can be used for  the verification of the
hypotheses used for the calculation of corrections to this ansatz. It is believed (but not yet proved)  that the  true   amplitudes can be presented as the product of $M_{BDS}$ and  the remainder function $R$, where  $M^{BDS}$ contains all infrared divergences  and
$R$ depends only on the anharmonic ratios of kinematic
invariants  \cite{Bern:2006ew}-\cite{Nguyen:2007ya}.
This property is called dual conformal
invariance.  Another property is the  conjecture (also not yet proved) of correspondence  between the MHV amplitudes and expectation values of Wilson loops \cite{Drummond:2007aua, Drummond:2007cf}, \cite{Brandhuber:2007yx}-\cite{Drummond:2008aq}.  All this makes important the direct calculation of the discontinuities.

The paper is organized as follows. In
the next Section we  introduce the notation, give the general expression for the discontinuities and use it for the calculation of  the discontinuity of the amplitude ${\cal A}^{\{-\}}_{2\rightarrow 2}$. Discontinuities of the amplitude ${\cal A}^{\{-\}}_{2\rightarrow 3}$ are found in Section 3. Section 4 is devoted to the calculation of discontinuities  of the amplitude ${\cal A}^{\{-\}}_{2\rightarrow 4}$. Discontinuities of amplitudes with a larger number of final particles are considered in Section 5. Conclusions
are drawn in Section 6. Appendices A, B  and C contain some details of calculations.

\section{Definitions, notation and the ${\cal A}^{\{-\}}_{2\rightarrow 2}$ discontinuity}

Let us first present explicit forms  of the  gluon trajectory and  the Reggeon  vertices in $N=4$ SYM with the accuracy up to terms vanishing in the limit $\epsilon \rightarrow 0 $. In the NLO the vertices, as well as the impact factors,  are scheme-dependent. We will use the scheme introduced in Ref.~\cite{Fadin:1998fv} and   then developed in Refs.~\cite{Fadin:2006bj},
which we call  standard one.  But since usual dimensional regularization is incompatible with supersymmetry,  we will  use  the dimensional reduction instead of the dimensional regularisation.  The NLO trajectory is given by
\cite{Fadin:1995dd}-\cite{Gerasimov:2010zzb}
\be
\omega(t) = -2\bar g^2\left(\frac{1}{\epsilon} +\ln(-t)\right)+2\bar g^4\left[\zeta(2)\left(\frac{1}{\epsilon} + 2\ln(-t)\right)-\zeta(3)\right]
 ~,  \label{omega}
\ee
where $\zeta(n)$ is the Riemann zeta-function,
\be
\bar{g}^{2}
=\frac{g^{2}N_{c}\Gamma(1-\epsilon)}{(4\pi)^{2+{\epsilon}}}~,~~~~\; \epsilon=\frac{D-4}{2}~,
\ee
$\Gamma(x)$ being the Euler  gamma-function and $D$ is the space-time dimension.

For the gluon polarization vectors in the Reggeon vertices and impact factors we will use  the $L$ and $R$ light-cone gauges $(e^Ln_2)=0$ and $(e^Rn_1)=0$ respectively,  with the light-cone vectors $n_2$ and $n_1$
such that
\be
(n_1n_2) =1, \; \, (p_Ap_B)  \simeq  (p_A n_2)(p_B n_1)~.
\ee
Then,
\begin{equation}
e^L=e^L_{\perp}-\frac{(e^L_{\perp}k_{\perp})}{kn_2}n_2\,, \qquad
\,e^R=e^R_{\perp}-\frac{(e^R_{\perp}k_{\perp})}{kn_1}n_1\,. \label{axial gauges}
\end{equation}
Note that the transverse parts of the polarization vectors in the left and right gauges are different.  It is easy to see that the polarization
vectors are connected by the  gauge transformation:
\begin{equation}
e^L = e^R-2\frac{(e^R_{\perp}k_{\perp})}{k^{2}_{\perp}}k~, ~~~~\;\; e^R = e^L-2\frac{(e^L_{\perp}k_{\perp})}{k^{2}_{\perp}}k~.  \label{gauge transformation}
\end{equation}
For transverse components this means
\begin{equation}
e^L_{\perp \mu}\ = \Omega_{\mu\nu}e^{R\; \nu}_{\perp }\,, \qquad e^{R}_{\perp \mu}\
= \ \Omega_{\mu\nu} e^{L\; \nu}_{\perp }\,, \label{gauge
transformation of e perp}
\end{equation}
where
\be
\Omega_{\mu\nu}\ =\ \Omega_{\nu\mu}\ =\
g^{\perp}_{\mu\nu}-2\frac{k_{\perp\mu}k_{\perp\nu}} {k_{\perp}^2}\,,  \qquad
\Omega_{\mu\nu}\Omega^{\nu\rho}\ =\ g_{\mu}^{\rho}\,. \label{gauge transformation
Omega}
\ee
Using the results of  Refs.~\cite{Fadin:1995xg,Fadin:1993wh},
\cite{Fadin:1992zt} and \cite{Kozlov:2014gaa} for the  one-loop gluon, quark and scalar corrections correspondingly, for the gluon-gluon-Reggeon vertex we have
\be
\Gamma^R_{G'G} = g T^R_{G'G} (\vec e^{\;*\prime}\vec e^{\;})\left[ 1+\bar g^ {\;2}(\qs)^\epsilon \left(-\frac{2}{\epsilon^2} +5\zeta(2)\right)\right]
.\;
\ee
Here $q$ is the Reggeon momentum, $\vec e$ and $\vec e^{\;*\prime}$ are the polarisation vectors  of the initial and final gluons $G$ and  $G'$ respectively (they have to be taken  in the same gauge), $T^R_{G'G}$ is the colour group generator in the adjoint representation.  For simplicity, here and in the following we use for  colour indices the same letters as  for particles and Reggeons.

The Reggeon-Reggeon-gluon vertex   was obtained  in the Born approximation in  Ref.~\cite{Lipatov:1976zz} and looks as
\begin{equation}
\gamma_{R_1R_{2}}^{G(B)}\ =\ gT_{R_1R_{2}}^{G}e^*_{\mu}(k)C^\mu (q_2,q_1)\,,
\label{gamma-RRG}
\end{equation}
where
\begin{eqnarray}
&& \hspace*{-1cm} C_{\mu}(q_2,q_1)\ =\ -q_{1\mu}-q_{2\mu}+p_{1\mu}
\left(\frac{q_1^2}{kp_1}+2\frac{kp_2}{p_1p_2}\right) -
p_{2\mu}\left(\frac{q_2^2}{kp_2}+2\frac{kp_1}{p_1p_2}\right)
\nonumber\\
&& =\ -q_{1\perp\mu}-q_{2\perp\mu}-\frac{p_{1\mu}}{2(kp_1)}
\left(k_{\perp}^2-2q_{1\perp}^2\right)+\frac{p_{2\mu}}{2(kp_2)}
\left(k_{\perp}^2-2q_{2\perp}^2\right).
  \label{vector-C}
\end{eqnarray}
The vertex is gauge invariant, being $C_{\mu}(q_2,q_1)k_\mu=0$.
In the light cone gauges \eqref{axial gauges} we get
\be
e_\mu^*(k)C_\mu(q_2,q_1)\ =\ e^{L*}_\perp C^L_\perp(q_2,q_1) =  e^{R*}_\perp C^R_\perp(q_2,q_1)~, \; \;
\ee
where
\[
C^L_\perp(q_2,q_1)  =C_\perp(q_2,q_1) -\frac{n_2 C(q_2,q_1)}{kn_2}k_\perp = -2\left(q_{1\perp} -k_{\perp}\frac{q_{1\perp}^2}{k_{\perp}^2}\right),\;\;
\]
\begin{equation}
C^R_\perp(q_2,q_1)  =C_\perp(q_2,q_1) -\frac{n_1 C(q_2,q_1)}{kn_1}k_\perp = -2\left(q_{2\perp} +k_{\perp}\frac{q_{2\perp}^2}{k_{\perp}^2}\right).\;\;
\label{vector C L R}
\end{equation}

It makes sense to note that using the light-cone gauges does not mean loss of generality.  One  can restore  any vertex in a gauge  invariant  form from its form in one of the gauges \eqref{axial gauges}.  Let us demonstrate it here for the vertex \eqref{gamma-RRG}, denoting $C(q_2,q_1)$ there as $C$  for brevity. Note that $C$  can be changed by adding terms  proportional
to $k$  without  changing  the vertex \eqref{gamma-RRG}, as well as $ C^L_\perp$  and  $ C^R_\perp$ defined in  formulas \eqref{vector C L R}, and without loss of the gauge invariance. Let us choose  these terms in such a way that $C$  goes to $C^L$  subject to the condition $(C^L n_2)=0$. Then, we have
\be
C^L= C^L_\perp + \frac{n_1C^L}{n_1n_2}n_2~.
\ee
On the other hand, from  $kC^L=0$ we have
\be
C^L_\perp k_\perp +\frac{(n_1 C^L)(n_2 k)}{n_1n_2} =0~,
\ee
so that
\be
C^L=C_\perp^L-\frac{C_\perp^L k_\perp}{kn_2}n_2~.
\ee
As  it has been said, using in  Eq.~\eqref{gamma-RRG} $C^L$ instead of  $C$ does not change the vertex leaving it gauge invariant. Thus, we obtain the gauge-invariant form  of the vertex from its form in the light-cone gauge. Using the relations \eqref{vector C L R}
one can see  that  $C^L$ is equal to $C -k$, where $C$ is the original form given by Eq.~\eqref{vector-C}.

One-loop gluon corrections to the vertex were calculated  in Refs. \cite{Fadin:1993wh}, \cite{Fadin:1996yv}-\cite{Fadin:2000yp}. In the last paper they were  obtained at arbitrary $D=4+2\epsilon$ dimension.  With the same accuracy, the  quark  and scalar corrections  were obtained in Refs.~\cite{Fadin:1994fj} and  \cite{Gerasimov:2010zzb} respectively. In the $N=4$ SYM,  with the accuracy resulting when  the  terms singular at small  $\vec k$  are given  at arbitrary D, but the other terms in the limit $\epsilon \rightarrow 0$, we have in the dimensional reduction
\begin{equation}
\gamma_{R_1R_2}^{G}(q_1,q_2) =\gamma_{R_1R_2}^{G(B)}(q_1,q_2)\biggl(1-\bar{g}^2\biggl[ \frac{(\vks)^\epsilon}{\epsilon^2}-\frac{\pi^2}{2}+ \frac{1}{2}\ln^2\Big(\frac{\qs_{1}}{\qs_{2}}\Big)\biggr]\biggr)~.
\label{vertex}
\end{equation}

A general representation for the discontinuities was derived in Ref.~\cite{Fadin:2006bj}  (it is presented also in Ref.~\cite{Ioffe:2010zz}).
The discontinuity of ${\cal A}^{\{-\}}_{2\rightarrow n+2}$  in the $s_{i,j}$-channel is represented as
\[
-4i(2\pi)^{D-2}\delta\left(q_{i\bot}-q_{(j+1)\bot}-\sum_{l=i}^{l=j}
k_{l\bot}\right)\mathrm{disc}_{s_{i,j}}{\cal A}_{2\to n+2}\ =\
\frac{{\Gamma}^{R_1}_{A'A}}{t_{1}}\Big(\frac{s_1}{|\q_{1}||\vk_{1}|}\Big)^{\omega(t_1)}
\]
\[
\times\ \left(\prod_{l=2}^i
 \frac{\gamma^{G_{l-1}}_{R_{l-1}R_l}}{t_{l}} \Big(\frac{s_l}{|\vk_{l-1}||\vk_{l}|}\Big)^{\omega(t_l)}\right)
 \langle G_i R_i|\left( \prod_{l=i+1}^{j-1}
 \Big(\frac{s_l}{|\vk_{l-1}||\vk_{l}|}\Big)^{\hat{\cal K}}\hat{\cal G}_l\!\right)
\Big(\frac{s_j}{|\vk_{j-1}||\vk_{j}|}\Big)^{\hat{\cal K}} | G_j R_{j+1}\rangle
\]
\be
\times\
\left(\prod_{l=j+1}^n
  \Big(\frac{s_l}{|\vk_{l-1}||\vk_{l}|}\Big)^{\omega(t_l)} \frac{\gamma^{G_{l}}_{R_{l}R_{l+1}}}{t_{l}}\right)
 \Big(\frac{s_{n+1}}{|\vk_{n}||\q_{n+1}|}\Big)^{\omega(t_{n+1})}\frac{\Gamma^{R_{n+1}}_{B'B}}{t_{(n+1)}}\,. \label{disc s ij}
\ee
Here the bra- and ket-states states  $\langle G_iR_i|$ and $| G_j R_{j+1}\rangle$  denote   the impact factors for  the Reggeon-gluon   transitions,    $\widehat{\cal K}$    and $\widehat{\cal G}_l$  are  the operators of the BFKL kernel and the gluon  production, which  acts in the space of states $|{\cal G}_1{\cal G}_2\rangle$ of two $t$-channel Reggeons with the  orthonormality property
\be
\langle{\cal G}'_1{\cal G}'_2|{\cal G}_1{\cal G}_2\rangle =\xs_1\xs_2 \delta(\x_1-\xp_1)\delta(\x_2-\xp_2)\delta_{c_1c'_1}\delta_{c_2c'_2}~, \label{orthonormality}
\ee
where $\x_i$ and $\xp_i$ are the Reggeon transverse  momenta and $c_i$ and $c'_i$ are their colour indices. The operators are specified by their matrix elements and  the states are defined by their projections on the two-Reggeon states.

If  $i=0$ we must omit all factors to the left of $\langle G_0  R_0|$ and replace $\langle G_0 R_0|$ by the impact factors of  $A\rightarrow A'$ transition  $\langle A^{\prime} A|$ and   $k_0-q_0$ by $p_{A'}-p_A$; in the
case $j=n+1$ we must omit all factors to the right of $| G_{n+1} R_{n+2}\rangle$ and
perform the substitutions $|G J_{n+1} R_{n+2}\rangle \rightarrow | B^{\prime} B\rangle, \;\; k_{n+1}+q_{n+2}\rightarrow p_{B'}-p_B$.

For the discontinuity $ disc_s{\cal A}_{2\rightarrow 2}  ={\cal A}_{2\rightarrow 2}(s+i0) -{\cal A}_{2\rightarrow 2}(s-i0)$ we have
\be
-4i(2\pi)^{D-2}\delta(\q-\q_B)\,disc_s{\cal A}_{2\rightarrow 2} = 2s \langle A' A|e^{\hat {\cal K} \ln\left(\frac{s}{\qs}\right)}|B'B\rangle~,
\label{2-2 disc}
\ee
where  $\q =p_{A} -p_{A'},  \;\; \q_B =p_{B'} -p_B $\,.

We have to pay attention here on the fundamental difference between the sense of the representation   \eqref{2-2 disc} used here  and that of the formally quite similar  representation of the discontinuities of amplitudes with the Pomeron exchange.  The BFKL Pomeron means the positive signature and the colour singlet  in the $t$-channel, while the  amplitudes considered in this paper  are  the amplitudes with the negative signature and the adjoint representation of the colour group in the $t$-channel.  The gluon Reggeization makes the discontinuities  \eqref{2-2 disc} much simpler  than the discontinuities of amplitudes with the Pomeron exchange. Indeed, the bootstrap conditions of  the gluon Reggeization \cite{Fadin:2006bj} tells us that
\be
\langle A' A| = \Gamma^R_{A' A}g\langle R_\omega(\q)|~,~~~~\;\; |B'B\rangle = g| R_\omega(\q_B)\rangle \Gamma^R_{B' B}~, \;\; \label{if bootstrap}
\ee
\be
{\hat {\cal K}}|R_\omega(\q)\rangle =\omega(t)|R_\omega(\q)\rangle~, \;\;  \label{s.f. bootstrap}
\ee
\be
  \frac{g^2\qs}{2{(2\pi)^{D-1}}}\langle R'_\omega(\qp)|R_\omega(\q)\rangle =
\delta_{R'R}\delta(\q-\qp)\omega(t)~, \;\;  \label{normalisation bootstrap}
\ee
where $\Gamma^R_{A' A}$ and $\Gamma^R_{B' B}$ are the scattering Reggeon vertices entering in the form \eqref{A 2-2+n}, $| R_\omega(\q)\rangle $ is the process independent eigenstate of the kernel ${\hat {\cal K}}$ with eigenvalue $\omega(t)$ and  normalization \eqref{normalisation bootstrap}. It is transformed according to the adjoint representation of the colour group.  In the right side of Eq.~\eqref{normalisation bootstrap}  $R'$ and $R$ are the colour indices of the eigenstates; in Eq.~\eqref{if bootstrap} summation over the colour indices $R$ is assumed.  Note that the bra and ket vectors are related by the  left-right substitution, where $A\leftrightarrow B, \; A'\leftrightarrow B', \; n_1\leftrightarrow n_2$, that means, in particular, replacement of the left and right gauges.

Fulfilment of the bootstrap conditions  \eqref{if bootstrap}-\eqref{normalisation bootstrap} was  proved in the NLO both in  QCD  \cite{Fadin:2000ww, Fadin:2002hz} and SYM \cite{Kozlov:2014gaa}.
Using these conditions we have from the representation \eqref{2-2 disc}
\be
disc_s{\cal A}_{2\rightarrow 2} = -i\pi\left(\frac{2s}{t}\right) \omega(t)\Gamma^R_{A' A} \left(\frac{s}{\qs}\right)^{\omega(t)} \Gamma^R_{B' B} ~.
\label{2-2 disc reduced}
\ee
It is easy to see that the result \eqref{2-2 disc reduced} with account of the form \eqref{A 2-2+n}
at $n=0$  is in agreement with the bootstrap relation \eqref{bootstrap relations}.

Finally, let us present the eigenstate $|R_{\omega}(\q)\rangle$. From Refs.~\cite{Fadin:2000ww} (see also Ref.~\cite{Kozlov:2011zza}) and  \cite{Kozlov:2013zza} we obtain with the accuracy up to terms vanishing in the limit $\epsilon \rightarrow 0 $
\be
\langle{\cal G}_1{\cal G}_2|R_{\omega}(\q)\rangle  =\delta(\q-\x_1-\x_2)T^R_{{\cal G}_1{\cal G}_2} \left(1+ \bar g ^2\left[-\zeta(2) -\frac12 \ln\left(\frac{\xs_1}{\qs}\right)\ln\left(\frac{\xs_2}{\qs}\right)\right]\right)~, \label{R omega}
\ee
where $\x_1$ and $\x_2$ are the momenta of the Reggeons ${\cal G}_1$  and ${\cal G}_2$ respectively. 

It is necessary to note here that the accuracy of Eq. \eqref{R omega} does not provide preservation of nonvanishing  in the limit $\epsilon \rightarrow 0 $ terms of the  $\bar g ^2$ order in the product 
 \[
\langle R'_\omega(\qp)|R_\omega(\q)\rangle =\sum_{{\cal G}_1{\cal G}_2}\int \langle R'_\omega(\qp)|{\cal G}_1{\cal G}_2\rangle \frac{d\x_1d\x_2}{\xs_1\xs_2} \delta(\q-\x_1-\x_2)\langle{\cal G}_1{\cal G}_2|R_\omega(\q)\rangle
\]
(the summation here is performed over colour states of the Reggeons ${\cal G}_1$ 
and ${\cal G}_2$) because of the infrared divergency of the integration measure. 
To provide the preservation one has  to keep in $\langle{\cal G}_1{\cal G}_2|R_{\omega}(\q)\rangle $ terms of the order ${\cal O} (\bar g ^2\epsilon)$. 

\section{Discontinuites of the $2\rightarrow 3$ amplitude} 
\subsection{Discontinuites in the $s_1$ and $s_2$ channels}    

For the  $s_1$-channel discontinuity we obtain  from the general form \eqref{disc s ij}
\be
-4i(2\pi)^{D-2}\delta(\q_1-\vk -\q_2)\,disc_{s_1}{\cal A}_{2\rightarrow 3} = 2s \langle A' A|e^{\hat {\cal K} \ln\left(\frac{s_1}{|\q_1||k_{1}|}\right)}|GR_2\rangle \frac{1}{t_{2}}\Big(\frac{s_{2}}{|k_{1}||\q_{2}|}\Big)^{\omega(t_{2})}
\Gamma^{{R}_{2}}_{B'B}~, \label{s 1 2-3 disc}
\ee
where $q_1=p_A-p_{A'}, \; k$ and $q_2$ are the momenta of the gluon $G$ and the Reggeon $R_2$, $|GR_2\rangle$  is  the impact factor for the  Reggeon-gluon transition.
The bootstrap conditions \eqref{if bootstrap} and \eqref{s.f. bootstrap} give us
\[
-4i(2\pi)^{D-2}\delta(\q_1-\vk -\q_2)\,disc_{s_1}{\cal A}_{2\rightarrow 3} = 2s \Gamma^{R}_{A' A}\Big(\frac{s_{1}}{|k_{1}||\q_{1}|}\Big)^{\omega(t_{1})} \frac{1}{t_{2}}\Big(\frac{s_{2}}{|k_{1}||\q_{2}|}\Big)^{\omega(t_{2})}
\Gamma^{{R}_{2}}_{B'B}
\]
\be
\times g\langle R_{\omega}(\q_1)|GR_2\rangle \label{s 1 2-3 disc 1}~.
\ee
The impact factors for Reggeon-gluon transitions were calculated in  Refs.~\cite{Kozlov:2012zz, Kozlov:2013zza} in the special scheme (the so called bootstrap scheme) which simplifies the proof of  the most complicated  bootstrap condition
\be
\langle G R_1| -g\qs_1\langle R_{\omega}(\q_1)|\hat{{\cal G}} =g\gamma^G_{R_1R}
\langle R_{\omega}(\q_1-\vk)|~,
\label{G bootstrap}
\ee
where  $\hat{{\cal G}}$ is the gluon  production operator, $k$ is the gluon momentum.     The Reggeon $R$ in the condition \eqref{G bootstrap} has the momentum $q_1-k$ and the same colour indices as the eigenstate  $\langle R_{\omega}(\q_1-\vk)|$; summation over them is assumed.
The eigenfunction $\langle R_{\omega}(\q_1)|{\cal G}_1{\cal G}_2\rangle$ in the bootstrap  scheme also was obtained in Ref.~\cite{Kozlov:2013zza}. We could calculate the matrix element $\langle R_{\omega}(\q_1)|GR_1\rangle$ in  Eq.~\eqref{s 1 2-3 disc 1} just in this scheme.  It turns, however, that it is much more convenient, especially in the further calculations,  to use the  scheme  which we call conformal. It is associated with the modified kernel $\hat {\cal K}_m$, introduced in Ref.~\cite{Bartels:2008sc}, which is obtained from the usual BFKL kernel in the adjoint representation, by subtraction of the gluon trajectory depending on the total $t$-channel momentum.  One of advantages of this  kernel is its infrared safety, which permits to consider this kernel at  physical transverse dimension $D-2 =2$.  But the most important advantage is its behaviour under  M\"{o}bius  transformations in the two-dimensional transverse momentum space. It is not difficult to see that in the leading order ${\cal K}_m$ is M\"{o}bius  invariant.  But in the NLO in the standard scheme, in which the kernel was initially  calculated \cite{Fadin:2000kx,Fadin:2000hu}, it is not  M\"{o}bius invariant.  The existence of the scheme where the modified kernel is   M\"{o}bius invariant (M\"{o}bius scheme) was conjectured in  Ref.~\cite{Fadin:2011we} and then proved  in Ref.~\cite{Fadin:2013hpa}, where the transformation from the standard ${\cal K}_m$ to  the conformal (M\"{o}bius invariant) kernel ${\cal K}_c$ was found. It reads
\be
{\hat{\cal K}}_c = {\hat{\cal K}}_m - \frac14\left[{\hat{\cal K}}^B\left[ \ln\left({\hat{\vec q}_1^{\,2}}{\hat{\vec q}_2^{\,2}}\right),{\hat{\cal K}}^B\right]\right]~, \label{kernel transformation}
\ee
where ${\hat{\cal K}}^B$ is the LO kernel. Note that since ${\hat{\cal K}}$ and ${\hat{\cal K}}_m$ differ only for the trajectory depending on the total $t$-channel momentum, which is a $C$-number, in all commutators  ${\hat{\cal K}}$ can be replaced by ${\hat{\cal K}}_m$ and vice versa. We will use the following representations for the kernel:
\be
\langle{\cal G}'_1{\cal G}'_2|{\hat{\cal K}}|{\cal G}_1{\cal G}_2\rangle =\delta(\x_1+\x_2-\xp_1-\xp_2)\ \qs \sum_R ({\cal P}_R)_{{\cal G}_1{\cal G}_2}^{{\cal G}'_1{\cal G}'_2}
\; K^R (\x_1,\x_2; \, \vl )~. \label{K representation}
\ee
Here $\x_i$ and $\xp_i$ are the Reggeon momenta, $\q= \x_1+\x_2, \;\; \vl =\x_1-\xp_1$, ${\cal P}_{R}$ is the projection operator on the representation $R$ of the colour group,
and
\be
K^R(\x_1,\x_2; \, \vl ) = K^R_r(\x_1,\x_2; \, \vl ) + \frac{\xs_1\xs_2}{\qs}\left(\omega(-\xs_1)\delta(\x_1-\xp_1)+\omega(-\xs_2)\delta(\x_2-\xp_2)\right)~, \label{K = K r + omega}
\ee
where $K^R_r$ is called the real part of the kernel.

In general, the kernel $K^R_r(\x_1,\x_2; \, \vl )$   depends on $R$. But at large $N_c$ only the antisymmetric and symmetric adjoint  representations do survive in the decomposition \eqref{K representation}, with
\be
\left({\cal P}_{Aa}\right)_{{\cal G}_1{\cal G}_2} ^{{\cal G}'_1{\cal G}'_2}= \frac{1}{N_c}f_{i{\cal G}_1{\cal G}_2}f_{i{\cal G}'_1{\cal G}'_2}~, ~~~~\;\; \left({\cal P}_{As}\right)_{{\cal G}_1{\cal G}_2} ^{{\cal G}'_1{\cal G}'_2}= \frac{1}{N_c}d_{i{\cal G}_1{\cal G}_2}d_{i{\cal G}'_1{\cal G}'_2}~,
\ee
and the same kernel $K^R_r(\x_1,\x_2; \, \vl )$. Therefore in the following we will omit the index of representation $R$.

In the LO the real part of the kernel is given by
\be
K^B_r(\x_1,\x_2; \, \vl )=\frac{g^2\ N_c}{2(2\pi)^{D-1}}\left(\frac{\xs_1\xps_2+\xs_2\xps_1}{\qs\vls} -1\right)~, \label{K B r}
\ee
whereas the gluon trajectory has the representation
\be
\omega^B(t) = \frac{g^2\ N_c\  t}{2(2\pi)^{D-1}}\int\frac{d\l}{\vls(\q-\vl)^2}~,~~~~\; \;t=-\qs~.
\label{omega B}
\ee

The difference with usual denotation is in the factor $\qs $ in the representation \eqref{K representation}.
Its extraction is necessary to make the modified  kernel
\be
K_m(\x_1,\x_2; \, \vl )= K(\x_1,\x_2; \, \vl )-\frac{\xs_1\xs_2}{\qs}\delta(\x_1-\xp_1)\delta(\x_2-\xp_2)\;\omega(t)
\ee
explicitly invariant at $D=4$  with respect to the M\"{o}bius transformations
\be
z_i\rightarrow \frac{az_i+b}{cz_i+d_i}~,
\ee
where $a, \; b, \; c$ and $\; d$ are complex numbers, $z_i=x_i+iy_i,  \; $   $x_i$ and $y_i$ are the Cartesian components  of the  ``dual" transverse momenta $\p_i$ such that
\be
\x_1=\p_1-\p_2, \;\; \x_2=\p_4-\p_1, \;\; \xp_1=\p_3-\p_2, \;\; \xp_2=\p_4-\p_3~.
\ee
The need for  the factor $\qs$ is clear from another point of view: the kernel could  be explicitly M\"{o}bius invariant only when the corresponding normalisation condition is   M\"{o}bius invariant. The condition  \eqref{orthonormality} is not invariant; to make it invariant one needs to multiply both sides  on $1/\qs$ and include  $1/\qs$ in the left-hand side in definition of the two-Reggeon states. This can be seen from the invariance of  the  corresponding measure,
\be
d\xp_1d\xp_2\frac{\qs }{\xps_1\xps_2}\delta(\x_1+\x_2 -\xp_1-\xp_2) = \frac{d^2z}{|z|^2}~,
\label{measure}
\ee
where $\q =\x_1+\x_2$ and $z=r^{'+}_2r^+_1/(r^{'+}_1r^+_2)$ is invariant.
Here and in the following  we use the chiral components  $r^+=x+iy$ and  $r^-=x-iy$  for the two-dimensional vectors $\vec r= (x,  y)$. Vice versa,  the two conjugate complex numbers $z$ and $z^*$ are confronted  with
the vector $\vec{z}$ through the components  $(z+z^*)/2$  and $(z-z^*)/(2i)$. At the same time, $d\vec r =dxdy =dr^+dr^-/2\,,\;\;
\delta(\x)=2\delta(r^+)\delta(r^-)\, $ and  we define $\delta^2(z)$ in such a way that $\delta^2(z) =\delta(z^+)\delta(z^-)/2 =\delta(\vec{z})$.

The transformation \eqref{kernel transformation} gives \cite{Fadin:2013hpa}
$K_c(\x_1,\x_2; \, \vl )=  K_c(z)$,  where $z=r_1^+r_2^{'+}/(r_2^+r_1^{'+})$
and
 \[
 K_c(z)=K_c^{B}(z)\left(1-\frac{g^2 N_c}{8\pi^2}\zeta(2)\right)+\delta^{(2)}(1-z)
 \left(\frac{g^2 \,N_{c}}{8\pi^{2}}\right)^2 \,3\zeta(3)
 +\frac{1}{8\pi}\left(\frac{g^2 \,N_{c}}{8\pi^{2}}\right)^2
 \]
 \[
 \times\Biggl[\left(\frac{1}{2}-\frac{1+|z|^2}{|1-z|^2}\right)\ln ^2{|z|^2}-\frac{1-|z|^2}{2|1-z|^2}\,
 \ln {|z|^2}\,\ln \frac{|1-z|^4}{|z|^2}
 \]
 \be
 +\left(\frac{1}{1-z}-\frac{1}{1-z^*}\right)(z-z^*)\,\int_0^1\frac{dx}{|x-z|^2}\
 \ln \frac{|z|^2}{x^2}\Biggr]\,~.  \label{K conformal}
 \ee
 Here
 \be
  K_c^B(z) =  \frac{g^2N_c}{32\pi^3}\left(\frac{z+z^*}{|1-z|^2}
 -\delta^{(2)}(1-z)\int\frac{d\vec l}{|l|^2}\frac{l+l^*}{|1-l|^2}\right)\,~,
 \ee
with the properties
\be
K_c(z)=K_c(z^*)= K_c(1/z), \;\;K_c(0)= 0~. \label{K c = 0}
\ee
The transformation \eqref{kernel transformation} has to be  accompanied by the corresponding transformation of the impact factors  and the eigenstate $\langle R_{\omega}|$. The eigenstate $\langle R_{\omega}|_c$  which corresponds to the kernel ${\hat{\cal K}}_c $ \eqref{kernel transformation} is
\be
\langle R_{\omega}(\q)|_c  = \langle R_{\omega}(\q)| -\frac14\langle R_{\omega}(\q)|^B\left[ \ln\left({\hat{\vec r}_1^{\,2}}{\hat{\vec r}_2^{\,2}}\right),{\hat{\cal K}}_r^B\right]~,   \label{R transformation}
 \ee
where ${\hat{\cal K}}_r^B$  is the real part of the LO kernel \eqref{K B r},
$\hat{\vec r}_1$ and $\hat{\vec r}_2$ are the Reggeon momentum operators.
Using (see Appendix A for details)
\be
\langle R_{\omega}(\q)|^B\left[ \ln\left({\hat{\vec r}_1^{\,2}}{\hat{\vec r}_2^{\,2}}\right),{\hat{\cal K}}_r^B\right]|{\cal G}_1{\cal G}_2\rangle= -2\bar g ^2\delta(\q-\x_1-\x_2)T^R_{{\cal G}_1{\cal G}_2}  \ln\left(\frac{\xs_1}{\qs}\right)\ln\left(\frac{\xs_2}{\qs}\right)~,
\label{rel 1}
\ee
we obtain from Eq.~\eqref{R omega}
\be
\langle R_{\omega}(\q)|{\cal G}_1{\cal G}_2\rangle_c  =\delta(\q-\x_1-\x_2)T^R_{{\cal G}_1{\cal G}_2} \left(1- \bar g ^2 \zeta(2)\right)~. \label{R c omega}
\ee
Now turn to the impact factor $|G R_2\rangle$. The impact factor corresponding to the kernel ${\hat{\cal K}}_c $ \eqref{kernel transformation} is obtained from  $|G R_2\rangle$ in the standard scheme by the  transformation
\be
|G R_2\rangle \rightarrow   |G R_2\rangle  +  \frac14 \left[ \ln\left({\hat{\vec r}_1^{\,2}}{\hat{\vec r}_2^{\,2}}\right),{\hat{\cal K}}^B\right] |G R_2\rangle^B ~.  \label{GR transformation 1}
\ee
It was found, however, in Ref.~\cite{Fadin:2014gra} that  impact factors for Reggeon-gluon transitions acquire the most simple form in the scheme in which  not only the kernel, but also the energy evolution parameter are  conformal invariant.  Transition to this scheme, which  is called conformal scheme,  means the additional transformation for the impact factor$|G R_2\rangle$,
\be
|G R_2\rangle \rightarrow    |G R_2\rangle   - \frac12  \ln\left(\frac{{\vec q}_1^{\,2}}{{\vec q}_2^{\,2}}\right){\hat{\cal K}_m}^B |G R_2\rangle^B ~.  \label{GR transformation 2}
\ee
Together with  the transformation \eqref{GR transformation 1} it gives
\be
|G R_2\rangle \rightarrow   |G R_2\rangle_c = |G R_2\rangle   -  \frac14 \left[ \ln\left({\hat{\vec r}_1^{\,2}}{\hat{\vec r}_2^{\,2}}\right),{\hat{\cal K}_r}^B\right] |G R_2\rangle^B - \frac12  \ln\left(\frac{{\vec q}_1^{\,2}}{{\vec q}_2^{\,2}}\right){\hat{\cal K}_m}^B |G R_2\rangle^B ~.  \label{GR transformation 3}
\ee
Note that the  transformation \eqref{GR transformation 2} does not affect the matrix element $\langle R_{\omega}(\q_1)|GR_2\rangle$ because $\langle R_{\omega}|$ is the eigenstate of ${\cal K}_m$ with the eigenvalue equal to $0$.

For amplitudes with the negative signature, the impact factors are  antisymmetric with respect to the  ${\cal G}_1\leftrightarrow {\cal G}_2$ exchange. In fact, putting
\be
\langle G R_1| =\langle G R_1|_s - \langle G R_1|_u~, \label{IF = s - u}
\ee
we have
\be
\langle G R_1|{\cal G}_1{\cal G}_2\rangle_u = \langle G R_1|{\cal G}_2{\cal G}_1\rangle_s~. \label{IFs - IFu}
\ee
As it follows from Ref.~\cite{Fadin:2014gra}, in the conformal scheme,  the impact factors $\langle G R_1|{\cal G}_1{\cal G}_2\rangle_s$ of   gluons with  the  polarisation vectors
\be
\vec{e}_{\lambda}^{\;L} = \frac{1}{\sqrt 2}\left(\vec e_x + i\lambda\vec e_y\right)~,~~~~ \;\; \vec{e}_{\lambda}^{\;L\,*} = \frac{1}{\sqrt 2}\left(\vec e_x  -i\lambda\vec e_y\right)~ \label{e L pm}~
\ee
for helicities $\lambda =\pm 1$  have the form
\be
\langle G R_1|{\cal G}_1{\cal G}_2\rangle_s =\langle G R_1|{\cal G}_1{\cal G}_2\rangle_s^B \Biggl[1+ \bar g^2\left( I_{\lambda}(z)-\frac12 \ln^2\left(\frac{\qs_1}
  {\qs_2}\right)
 -\frac{(\vks)^\epsilon}{\epsilon^2}+2\zeta(2)\right)\Biggr]~. \label{left if}
\ee
Here $z=-q_1^+r_2^{+}/(k^+ r_1^{+})$,
\be
\langle G R_1|{\cal G}_1{\cal G}_2\rangle_s^B = -g^2\delta(\q_1-\x_1-\x_2-\vk)
\left(T^{R_1}T^G\right)_{{\cal G}_1{\cal G}_2} \vec{e}_{\lambda}^{\;L\,*} \vec{C}^{\;L}(\x_1, \q_1)~, \label{left if b}
\ee
\be
 \vec{C}^{\;L}(\x_1, \q_1)= -2\left(\q_1 - (\q_1-\x_1)\frac{\qs_1}{(\q_1-\x_1)^2}\right)~, \label{left C}
\ee
and  $I_{+1}(z) =I(z), \;\;  I_{-1}(z)=I^*(z) =I(z^*) $, where
\[
I(z)=\frac{1-z}{8}\Biggl(
\ln\biggl(\frac{|1-z|^2}{|z|^2}\biggr)\ln\biggl(\frac{|1-z|^4}{|z|^6}\biggr)
-6Li_2(z)+6Li_2(z^*) -3\ln|z|^2
\ln\frac{1-z}{1-z^*} \Biggr)
\]
\be
-\frac12\ln|1-z|^2\ln\frac{|1-z|^2}{|z|^2} -\frac38\ln^2|z|^2~,  \label{I +}
\ee
\begin{equation}
Li_2(z)=-\int_0^1\frac{dx}{x}\ln(1-xz)~.
\end{equation}
Note that $I(0)=0, \;\;  I(1/z)=I(z)/z$.
In the two dimensional transverse momentum space,  with the polarization vectors \eqref{e L pm} we have
\be
\vec{e}_{+}^{\;L\,*}\vec{C}^{\;L}(\x_1, \q_1)  =\sqrt 2 \frac{q_1^-r_1^+}{(q_1-r_1)^+}~,~~~~ \;\;
\vec{e}_{-}^{\;L\,*}\vec{C}^{\;L}(\x_1, \q_1) =\frac{q_1^+r_1^-}{(q_1-r_1)^-}~.
\label{(e L pm C L)}
\ee

The set of diagrams for the process   $A+B\rightarrow A' +G +B'$ is evidently invariant with respect to rotating around the gluon line and the exchange $A\leftrightarrow B$. It means that the impact factor $\langle {\cal G}_1{\cal G}_2 |G R_2 \rangle$ can be obtained  from  $\langle G R_1|{\cal G}_1{\cal G}_2\rangle$ by the replacement
\be
n_1\leftrightarrow n_2, ~~\;\; \q_1\rightarrow -\q_2, ~~\;\; \x_{1,2}\rightarrow -\x_{1,2}~, ~~\;\; \left(T^{R_1}T^G\right)_{{\cal G}_1{\cal G}_2} \rightarrow \left(T^{R_2}T^G\right)_{{\cal G}_1{\cal G}_2}~. \label{L-R replacement}
\ee
The replacement $n_1\leftrightarrow n_2$ means also $\vec{e}_{\lambda}^{\;L} \leftrightarrow  \vec{e}_{\lambda}^{\;R}$.
With account of Eqs.~\eqref{gauge transformation of e perp} and \eqref{gauge transformation
Omega} we have
\be
\vec{e}_{\lambda}^{\;R} =-\left(\frac{k^+}{k^-}\right)^\lambda \vec{e}_{-\lambda}^{\;L}=-\left(\frac{k^+}{k^-}\right)^\lambda\left(\vec e_x  -i\lambda \vec e_y\right)~.
\label{polarization}
\ee
Using the  substitutions  \eqref{L-R replacement}  and formulas \eqref{polarization} we obtain
\be
\langle {\cal G}_1{\cal G}_2|G R_2\rangle_s =\langle {\cal G}_1{\cal G}_2|G R_2\rangle_s^B \biggl[1+ \bar g^2\left( I_\lambda^*(z)-\frac12 \ln^2\left(\frac{\qs_2}
  {\qs_1}\right)
 -\frac{(\vks)^\epsilon}{\epsilon^2}+2\zeta(2)\right)\biggr],  \label{R IF}
\ee
where $z=q_2^+r_2^{+}/(k^+ r_1^{+})$~,
\be
\langle {\cal G}_1{\cal G}_2|G R_2\rangle_s^B = g^2\delta(\x_1+\x_2-\vk-\q_2)
\left(T^{R_2}T^G\right)_{{\cal G}_1{\cal G}_2} \vec{e}_{\lambda}^{\;R\,*}\vec{C}^{\;R}(\q_2, \x_1)~ \label{R IF B}
\ee
and
\be
 \vec{C}^{\;R}(\q_2, \x_1)= -2\left(\q_2 + (\x_1-\q_2)\frac{\qs_2}{(\x_1-\q_2)^2}\right)~. \label{vec C R}
\ee
Using now Eqs.~\eqref{R c omega} and \eqref{R IF}-\eqref{vec C R}
we arrive to
\[
\langle R_{\omega}(\q_1)|GR_2\rangle_s
 = \frac{g^2N_c}{2 \qs_1 }\delta(\q_1-\vk-\q_2)T^G_{R_1R_2} \vec{e}_{\lambda}^{\;R\,*}\int d\x_1d\x_2\frac{\qs_1 }{\xs_1\xs_2}\delta(\x_1+\x_2 -\vk-\q_2)
\]
\be
\times\vec{C}^{\;R}(\q_2, \x_1)\biggl[1+ \bar g^2\left( I_{-\lambda}(z)-\frac12 \ln^2\left(\frac{\qs_2}
   {\qs_1}\right)
  -\frac{(\vks)^\epsilon}{\epsilon^2}+\zeta(2)\right)\biggr]~,  \label{R  GR}
\ee
where $\lambda = \pm 1$ is the gluon helicity,  $I_{+1}(z)=I (z), \; I_{-1}(z))=I(z^*)$,  $\; I (z)$ is defined in Eq.~\eqref{I +},  and  $\;\; z=q_2^+r_2^+/(k^+r_1^+)$.
The integral  with  $I_{-\lambda}(z)$ in Eq.~\eqref{R  GR} is not singular and can be calculated in two-dimensional transverse momentum space. Using the measure \eqref{measure} and
formulas
\[
\vec{e}_{+}^{\;R\,*}\vec{C}^{\;R}(\q_2, \x_1)  =\sqrt 2 \frac{k^-}{k^+}\frac{q_2^+r_1^-}{(r_1-q_2)^-} =\sqrt 2 \frac{q_2^+q_1^-}{k^+(1-z^*)}~,
\]
\be
\vec{e}_{-}^{\;R\,*}\vec{C}^{\;R}(\q_2, \x_1) =\sqrt 2 \frac{k^+}{k^-}\frac{q_2^-r_1^+}{(r_1-q_2)^+} =\sqrt 2 \frac{q_2^-q_1^+}{k^-(1-z)}~,  \label{(e R C)}
\ee
we obtain that the contribution of the term with $I_{-\lambda}(z)$ in Eq.~\eqref{R  GR} is equal zero. Indeed, for the positive helicity it is proportional to
\be
\int \frac{d^2z}{|z|^2(1-z^*)} I(z^*) =0~. \label{int I  = 0}
\ee
The result \eqref{int I  = 0} follows from the fact that  in the expansion of the integrand in powers of $(z^*)^n$ at  $|z|<1$  and in  powers of $(1/z^*)^n$ at  $|z|>1$ there are only terms with $n>0$ (remind that, as pointed out previously,  $I(z)=0, \;\; I(z) =zI(1/z)$). For the negative  helicity  the  result is obtained by complex conjugation.  It means that the term with $I_{-\lambda}(z)$ in Eq.~\eqref{R  GR}  can be omitted. The remaining integral (details of  the calculation are given in Appendix B)  is
\[
\int d\x_1d\x_2\frac{\qs_1 }{\xs_1\xs_2}\delta(\x_1+\x_2 -\vk-\q_2) \left(\q_2 + (\x_1-\q_2)\frac{\qs_2}{(\x_1-\q_2)^2}\right)
\]
\be
=\pi^{1+\epsilon}\Gamma(1-\epsilon)\left(\q_2+\vk\frac{\qs_2}{\vks} \right)
 \left(\frac{1}{\epsilon} +\ln\left(\frac{\qs_1\vks}{\qs_2}\right)\right)~. \label{int C}
\ee
For $disc_{s_1}{\cal A}^{\{-\}}_{2\rightarrow 3}/\Re {\cal A}_{2\rightarrow 3}$ from  Eqs.~\eqref{s 1 2-3 disc 1} and  \eqref{A 2-2+n} we get
\be
-4i(2\pi)^{D-2}\delta(\q_1-\vk-\q_2)\frac{disc_{s_1}{\cal A}^{\{-\}}_{2\rightarrow 3}}{\Re {\cal A}_{2\rightarrow 3}}
 =\frac{g\,t_1\langle R_{\omega}(\q_1)|GR_2\rangle}{\gamma^G_{R_1R_2}}
 =2\frac{g\,t_1\langle R_{\omega}(\q_1)|GR_2\rangle_s}{\gamma^G_{R_1R_2}}
~. \label{s 1 2-3 ratio}
\ee
Here the last equality  comes from antisymmetry of $\langle R_{\omega}| {\cal G}_1{\cal G}_2\rangle$ with respect to ${\cal G}_1\leftrightarrow {\cal G}_2$  exchange.
Then, using Eqs.~\eqref{vertex}, \eqref{gamma-RRG},  \eqref{R  GR} and  the equalities
\[
\vec{e}_{+}^{\;R\,*}\vec{C}^{\;R}(\q_2, \q_1) = \vec{e}_{+}^{\;L\,*}\vec{C}^{\;L}(\q_2, \q_1)  =\sqrt 2 \frac{q_2^+q_1^-}{k^+}~,  \label{(e C)}
\]
\be
\vec{e}_{-}^{\;R\,*}\vec{C}^{\;R}(\q_2, \q_1) = \vec{e}_{-}^{\;L\,*}\vec{C}^{\;L}(\q_2, \q_1) =\sqrt 2 \frac{q_2^-q_1^+}{k^-}~,
\ee
which means
\be
\gamma^{G (B)}_{R_1R_2}\Big|_{\lambda=+1}=- g T^G_{R_1R_2}\sqrt 2 \frac{q_2^+q_1^-}{k^+}, \;\; \gamma^{G (B)}_{R_1R_2}\Big|_{\lambda=-1}= -g T^G_{R_1R_2}\sqrt 2 \frac{q_2^-q_1^+}{k^-}~, \label{vertex helical}
\ee
we obtain
\be
\frac{disc_{s_1}{\cal A}^{\{-\}}_{2\rightarrow 3}}{\Re {\cal A}_{2\rightarrow 3}} =\pi i \bar g ^2\left(
\frac{1}{\epsilon} +\ln\left(\frac{\qs_1\vks}{\qs_2}\right)\right)\left(1-2\bar g^2\zeta(2)\right)~. \label{s_1 A23 final}
\ee
Here,  it is necessary to make the note analogous to that given at the end of Section 2. The accuracy of Eq. \eqref{R c omega}  for  $\langle R_{\omega}(\q)|{\cal G}_1{\cal G}_2\rangle_c $  and Eq. \eqref{R IF} for   $\langle {\cal G}_1{\cal G}_2|G R_2\rangle_s$ does not provide preservation of nonvanishing  in the limit $\epsilon \rightarrow 0 $ corrections of the  $\bar g ^2$ order in the integral \eqref{R  GR} (and therefore in the discontinuity $disc_{s_1}{\cal A}^{\{-\}}_{2\rightarrow 3}$)  because of the infrared divergency of the integration measure in  Eq.~\eqref{R  GR}. 
To provide the preservation one has  to find $\langle R_{\omega}(\q)|{\cal G}_1{\cal G}_2\rangle_c $  and    $\langle {\cal G}_1{\cal G}_2|G R_2\rangle_s$ 
with higher  accuracy. This issue requires special consideration. It  applies also to other discontinuities discussed below. 

Evidently, the $s_2$-channel discontinuity can be obtained by the replacement \eqref{L-R replacement} and is given by the relations
\be
-4i(2\pi)^{D-2}\delta(\q_1-\vk-\q_2)\frac{disc_{s_2}{\cal A}^{\{-\}}_{2\rightarrow 3}}{\Re {\cal A}_{2\rightarrow 3}}  =\frac{g \, t_2\langle GR_1|R_{\omega}(\q_2)\rangle}{\gamma^G_{R_1R_2}} \label{s_2 2-3 ratio}~,
\ee
\[
\frac{g\ t_2 \langle GR_1|R_{\omega}(\q_2)\rangle}{\gamma^G_{R_1R_2}} = \delta(\q_1-\vk-\q_2)
g^2\pi^{1+\epsilon}\Gamma(1-\epsilon)\left(1-2\bar g^2\zeta(2)\right)
\]
\be
\times\left(
\frac{1}{\epsilon} +\ln\left(\frac{\qs_1\vks}{\qs_2}\right)\right)~, \label{GR_1 R(q2)}
\ee
\be
\frac{disc_{s_2}{\cal A}^{\{-\}}_{2\rightarrow 3}}{\Re {\cal A}_{2\rightarrow 3}} =\pi i \bar g ^2\left(
\frac{1}{\epsilon} +\ln\left(\frac{\qs_2\vks}{\qs_1}\right)\right)\left(1-2\bar g^2\zeta(2)\right)~. \label{s_2 A23 final}
\ee

\subsection{Discontinuity in the $s$ channel}

According to the representation \eqref{disc s ij}, for the $s$-channel discontinuity we have
\be
-4i(2\pi)^{D-2}\delta(\q_1-\vk -\q_2)\,disc_{s}{\cal A}^{\{-\}}_{2\rightarrow 3} = 2s \langle A' A|e^{\hat {\cal K} \ln\left(\frac{s_1}{|\q_1||k|}\right)}\hat{{\cal G}}  e^{\hat {\cal K} \ln\left(\frac{s_1}{|k||\q_2|}\right)} |{B'B}\rangle~,    \label{s 2-3 disc}
\ee
where  $\hat{{\cal G}}$  is  the gluon production operator. Using the bootstrap conditions \eqref{if bootstrap} and    \eqref{s.f. bootstrap}, we obtain
\be
-4i(2\pi)^{D-2}\delta(\q_1-\vk-\q_2)\,)\frac{disc_{s}{\cal A}^{\{-\}}_{2\rightarrow 3}}{\Re {\cal A}_{2\rightarrow 3}}  = \frac{g^2\qs_1\qs_2\langle R_{\omega}(\q_1)|\hat{{\cal G}}|R_{\omega}(\q_2)\rangle}{\gamma^G_{R_1R_2}}~. \label{s  2-3 ratio}
\ee
Then, due to  the bootstrap condition \eqref{G bootstrap}, we have
\be
g\qs_1\langle R_{\omega}(\q_1)|\hat{{\cal G}}|R_{\omega}(\q_2)\rangle =
\langle G R_1|R_{\omega}(\q_2)\rangle  -g\gamma^G_{R_1R}
\langle R_{\omega}(\q_1-\vk)|R_{\omega}(\q_2)\rangle~.   \label{R G R}
\ee
Both matrix elements here are known: the first comes from  the calculation of the $s_2$-channel discontinuity, see Eq.~\eqref{GR_1 R(q2)},   and the second  from the bootstrap   condition  \eqref{normalisation bootstrap}. Thus, we obtain
\[
\frac{disc_{s}{\cal A}^{\{-\}}_{2\rightarrow 3}}{\Re {\cal A}_{2\rightarrow 3}} =-\pi i \bar g ^2\left(
\frac{1}{\epsilon} +\ln\left(\frac{\qs_2\vks}{\qs_1}\right)\right)\left(1-2\bar g^2\zeta(2)\right)- \pi i \omega(t_2)
\]
\be
=\pi i \bar g ^2\left[\frac{1}{\epsilon} +\ln\left(\frac{\qs_1\qs_2}{\vks}\right) +2
\bar g ^2 \left(\zeta(3)-\zeta(2)\ln\left(\frac{\qs_1\qs_2}{\vks}\right) \right) \right]~. \label{s A23 final}
\ee
In fact, it was not needed at all to calculate neither the $s_2$-channel, nor the  $s$-channel discontinuities, because they can be expressed in terms of $s$-channel discontinuities from the bootstrap relations \eqref{bootstrap relations}. Indeed, for   the amplitude ${\cal A}^{\{-\}}_{2\rightarrow 3}$ there are three relations:
\[
\frac{disc_{s_1}{\cal A}^{\{-\}}_{2\rightarrow 3}}{\Re {\cal A}_{2\rightarrow 3}}+\frac{disc_{s}{\cal A}^{\{-\}}_{2\rightarrow 3}}{\Re {\cal A}_{2\rightarrow 3}}  =-i\pi \omega(t_1)~,~~~~ \; \;  \frac{disc_{s_2}{\cal A}^{\{-\}}_{2\rightarrow 3}}{\Re {\cal A}_{2\rightarrow 3}}+\frac{disc_{s}{\cal A}^{\{-\}}_{2\rightarrow 3}}{\Re {\cal A}_{2\rightarrow 3}}  =-i\pi \omega(t_2)~,
\]
\be
\frac{disc_{s_1}{\cal A}^{\{-\}}_{2\rightarrow 3}}{\Re {\cal A}_{2\rightarrow 3}}-\frac{disc_{s_2}{\cal A}^{\{-\}}_{2\rightarrow 3}}{\Re {\cal A}_{2\rightarrow 3}}  =-i\pi \left(\omega(t_1) -\omega(t_2)\right) ~.
\label{2-3 relationships}
\ee
However, they are not independent:  the third of them is the difference of the first two.  Therefore, there are two relationships between the discontinuities, so that  only one of them is  independent. It is easy to see that
the discontinuities calculated above satisfy the relations \eqref{2-3 relationships}.  The fulfilment of the third of  them,  with account of
\be
\omega(t_1) - \omega(t_2) =2\bar g ^2\ln\left(\frac{\qs_2}{\qs_1}\right)\left(1-2\bar g ^2\zeta(2)\right)~,
\ee
follows  from Eqs.~\eqref{s_1 A23 final}  and  \eqref{s_2 A23 final} and
fulfilment of the second  follows from Eqs.~\eqref{s_2 A23 final} and \eqref{s A23 final}.

\section{Discontinuities of the $2\rightarrow 4$ amplitude}   
\subsection{Discontinuities in the $s_1$ and $s_3$ channels}   
From the representation \eqref{disc s ij}, for the $s_1$-channel discontinuity we have
\[
-4i(2\pi)^{D-2}\delta(\q_1-\vk_1 -\q_2)\,disc_{s_1}{\cal A}^{\{-\}}_{2\rightarrow 4} = 2s \langle A' A|e^{\hat {\cal K} \ln\left(\frac{s_1}{|\q_1||k|}\right)}|G_1R_2\rangle
\]
\be
\times\frac{1}{t_{2}}\Big(\frac{s_2}{|\vk_{1}||\vk_{2}|}\Big)^{\omega(t_2)}
\gamma^{G_2}_{R_2{R}_{3}}\frac{1}{t_{3}}\Big(\frac{s_{3}}{|k_{2}||\q_{3}|}\Big)^{\omega(t_{3})}
\Gamma^{{R}_{3}}
_{B' B}~,      \label{s 1 2-4 disc}
\ee
therefore, using the bootstrap relations \eqref{if bootstrap} and \eqref{s.f. bootstrap} and  the representation \eqref{A 2-2+n} of the MRK amplitude, we obtain
\be
-4i(2\pi)^{D-2}\delta(\q_1-\vk_1 -\q_2)\,\frac{disc_{s_1}{\cal A}^{\{-\}}_{2\rightarrow 4}}{\Re {\cal A}_{2\rightarrow 4}} =\frac{g^2\,t_1\,\langle R_{\omega}(\q_1)|G_1R_{2}\rangle}{\gamma^{G_1}_{R_1R_2}}~. \label{s 1 2-4 ratio}
\ee
The ratio in the right-hand side of Eq.~\eqref{s 1 2-4 ratio} is the same as in Eq.~\eqref{s 1 2-3 ratio} with the replacement $G\rightarrow G_1$, so that using Eq.~\eqref{s_1 A23 final}   we arrive to
\be
\frac{disc_{s_1}{\cal A}^{\{-\}}_{2\rightarrow 4}}{\Re {\cal A}_{2\rightarrow 4}} =\pi i \bar g ^2\left(
\frac{1}{\epsilon} +\ln\left(\frac{\qs_1\vks_1}{\qs_2}\right)\right)\left(1-2\bar g^2\zeta(2)\right)~. \label{s_1 A24 final}
\ee
Obviously, such ratio for the  $s_3$-channel discontinuity can be obtained by the replacement  $\vk_1\rightarrow \vk_2, \;\; \q_1\rightarrow -\q_3, \;\; \q_2\rightarrow -\q_2$; it reads
\be
\frac{disc_{s_3}{\cal A}^{\{-\}}_{2\rightarrow 4}}{\Re {\cal A}_{2\rightarrow 4}} =\pi i \bar g ^2\left(
\frac{1}{\epsilon} +\ln\left(\frac{\qs_3\vks_2}{\qs_2}\right)\right)\left(1-2\bar g^2\zeta(2)\right)~. \label{s_3 A24 final}
\ee
\subsection{Discontinuity in the $s_2$  channel}
For the $s_2$-channel discontinuity, using the modified kernel  $\hat {\cal K}_m$,  $\; \hat {\cal K} =\hat {\cal K}_m +\omega(t_2)$,  we have  from the representation \eqref{disc s ij}
\[
-4i(2\pi)^{D-2}\delta(\q-\q_B)\,disc_{s_2}{\cal A}^{\{-\}}_{2\rightarrow 4} = 2s\Gamma^{{R}_{1}}
_{A' A}\frac{1}{t_{1}}\Big(\frac{s_1}{|\q_{1}||\vk_{1}|}\Big)^{\omega(t_1)}\frac{1}{t_{3}}\Big(\frac{s_{3}}{|k_{2}||\q_{3}|}\Big)^{\omega(t_{3})}
\Gamma^{{R}_{3}}
_{B' B}
\]
\be
\times \Big(\frac{s_2}{|\vk_{1}||\vk_{2}|}\Big)^{\omega(t_2)}\langle G_1 R_1|e^{\hat {\cal K}_m \ln\left(\frac{s_2}{|\vk_1||\vk_2|}\right)}|G_2R_3)\rangle  ~.      \label{s 2 2-4 disc}
\ee
Here we meet two new important aspects. First, the energy dependence of the $s_2$ channel discontinuity  \eqref{s 2 2-4 disc} evidently differs from that predicted by the BDS ansatz \cite{Bern:2005iz}, where this dependence is the same as for the real part of ${\cal A}_{2\rightarrow 4}$.  Instead, according to
Eq.~\eqref{s 2 2-4 disc}, there is an additional dependence coming from the matrix element with $\hat {\cal K}_m$.
For agreement with the BDS ansatz the impact factors for Reggeon-gluon transitions have to be proportional to the eigenvector of $\hat {\cal K}_m$ with zero eigenvalue, what is obviously not so.  Note that the discrepancy
is manifested already in the leading logarithmic approximation.

Actually, it is well known that the BDS ansatz for $n$-gluon amplitudes is  incomplete at $n\geq 6$. The first indications of the incompleteness were obtained in Ref.~\cite{Alday:2007he} in the strong coupling regime using  the Maldacena hypothesis~\cite{Maldacena:1997re}  about the ADS/CFT duality,  and  in Ref.~\cite{Drummond:2007bm} using the hypothesis of  the scattering amplitude/Wilson loop correspondence.  Then  the incompleteness was shown by direct two-loop calculations in Ref.~\cite{Bern:2008ap}. Moreover, disagreement of the BDS ansatz with the BFKL approach is  also known~\cite{Bartels:2008ce}.  Dignity of the demonstration of the discrepancy  presented  here is its simplicity.

The second new aspect  is seen from the expressions for the impact factors in Eq.~\eqref{s 2 2-4 disc}
\[
\langle G_1 R_1|{\cal G}_1{\cal G}_2\rangle =\langle G_1 R_1|{\cal G}_1{\cal G}_2\rangle _s-\langle G_2 R_1|{\cal G}_2{\cal G}_1\rangle_s~, \;\;
\]
\be
\langle{\cal G}_1{\cal G}_2|G_2R_3\rangle =\langle{\cal G}_1{\cal G}_2|G_2R_3\rangle_s-\langle{\cal G}_2{\cal G}_1|G_2R_3\rangle_s~,
\ee
where $\langle G_1 R_1|{\cal G}_1{\cal G}_2\rangle_s$ is given by  Eqs.~\eqref{left if}--\eqref{I +} and $\langle{\cal G}_1{\cal G}_2|G_2R_3\rangle$ by Eqs.\eqref{R IF}--\eqref{vec C R} with the replacement $\vk\rightarrow \vk_2, \; \q_2\rightarrow \q_3$.   The new aspect  is the appearance in the discontinuity  of the colour structure $D^{G_1}_{R_1R_2}D^{G_2}_{R_2R_3}$, where $D^a_{bc} =d_{abc}$,  in addition to the structure $T^{G_1}_{R_1R_2}T^{G_2}_{R_2R_3}$ in the real part of the amplitude ${\cal A}_{2\rightarrow n+2}$ presented in  Eq.~\eqref{A 2-2+n}.  Indeed, using
\be
(T^aT^b)_{ij} f_{cij}=i\frac{N_c}{2}T^c_{ab}, \;\; (T^aT^b)_{ij}  d_{cij}=\frac{N_c}{2}D^c_{ab}~,
\ee
we have at large $N_c$
\[
(T^{R_1}T^{G_1})_{ij}(T^{R_3}T^{G_2})_{ij} =\frac{N_c}{4}\left(T^{G_1}T^{G_2} +D^{G_1}D^{G_2}\right)_{R_1R_3}~,\;\;
\]
\be
(T^{R_1}T^{G_1})_{ij}(T^{R_3}T^{G_2})_{ji} =\frac{N_c}{4}\left(-T^{G_1}T^{G_2} +D^{G_1}D^{G_2}\right)_{R_1R_3}~.
\ee

Writing explicitly all colour factors, we obtain using Eqs. \eqref{IF = s - u} and \eqref{IFs - IFu}
\[
\langle G_1 R_1|e^{\hat {\cal K}_m \ln\left(\frac{s_2}{|\vk_1||\vk_2|}\right)}|G_2R_3\rangle =\frac{N_c}{2}\left(T^{G_1}T^{G_2} +D^{G_1}D^{G_2}\right)_{R_1R_3} \langle \tilde{G_1 R_1}|_s \;e^{\hat {\cal K}_m \ln\left(\frac{s_2}{|\vk_1||\vk_2|}\right)}|\tilde{G_2R_3}\rangle_s
\]
\be
+\frac{N_c}{2}\left(-T^{G_1}T^{G_2} +D^{G_1}D^{G_2}\right)_{R_1R_3} \langle \tilde{G_1 R_1}|_s \;e^{\hat {\cal K}_m \ln\left(\frac{s_2}{|\vk_1||\vk_2|}\right)}|\tilde{G_2R_3}\rangle_u~, \label{decomposition of dis s 2}
\ee
where the tilde  sign  in the impact factors means rejection of colour factors.

It is very convenient  to use the  conformal representation for calculation of the matrix elements in the  right side of Eq.~\eqref{decomposition of dis s 2}, but transition to  the  two-dimensional transverse momentum space  in this representation   must be done with caution  because of the  infrared divergency in the first term  in the  right side of Eq.~\eqref{decomposition of dis s 2}. In the leading logarithmic approximation, this  problem  was considered in details in  Ref.~\cite{Bartels:2008sc}. In principle, nothing has changed at the transition to the next-to-leading approximation.

The  divergency  emerges because of the singularity of the integration measure \eqref{measure} at zero momenta of intermediate Reggeized gluons.  As it follows from Eqs. \eqref{left if}, \eqref{left if b}, \eqref{R IF} and \eqref{R IF B}, the $s$-pieces of the impact factors   $\langle G_1 R_1|{\cal G}_1{\cal G}_2\rangle_s$ and $\langle{\cal G}_1{\cal G}_2| G_2 R_3\rangle_s$ vanish at  $\x_1=0$, but not at $\x_2=0$. It means that the first matrix element in the  right side of Eq.~\eqref{decomposition of dis s 2} is divergent. Fortunately, the divergence exists only in the zero term of the expansion in powers of the BFKL kernel due to its property \eqref{K c = 0}. Therefore, writing
\[
\langle \tilde{G_1 R_1}|_s \;e^{\hat {\cal K}_m \ln\left(\frac{s_2}{|\vk_1||\vk_2|}\right)}|\tilde{G_2R_3}\rangle_s = \langle \tilde{G_1 R_1}|_s \;\left(e^{\hat {\cal K}_m \ln\left(\frac{s_2}{|\vk_1||\vk_2|}\right)}-1\right)|\tilde{G_2R_3}\rangle_s
\]
\be
+\langle \tilde{G_1 R_1}|_s |\tilde{G_2R_3}\rangle_s ~, \label{s s contribution}
\ee
we can use for the  first  term in the right side the conformal representation directly in the  two-dimensional  space.  Using  Eqs. \eqref{left if}, \eqref{left if b},
\eqref{(e L pm C L)} and \eqref{vertex}, \eqref{vertex helical} we have for the positive  helicity of the gluon $G_1$ ($\lambda_1=1$)
\be
\frac{\langle \tilde{G_1 R_1}|\tilde{{\cal G}_1{\cal G}_2}\rangle_s}{\tilde{\gamma}_{R_1R_2}} =g\delta(\q_1-\vk_1-\x_1-\x_2)\frac{1}{1-z_1} [1+\bar g^2 (I(z_1)-\zeta(2))]~, \label{left helical if}
\ee
where $ z_1=-q_1^+r_2^+/(k_1^+r_1^+)$ and the tilde signs means omission of the colour factors.  Analogously, for the positive  helicity of the gluon $G_2$ ($\lambda_2=1$), we obtain using  Eqs. \eqref{R IF}, \eqref{R IF B},
\eqref{(e R C)} and \eqref{vertex}, \eqref{vertex helical}
\be
\frac{\langle \tilde{{\cal G}_1{\cal G}_2}|\tilde{G_2 R_3}\rangle_s}{\tilde{\gamma}_{R_2R_3}} =-g\delta(\q_2-\vk_2-\x_1-\x_2) \frac{1}{1-z^*_2} [1+\bar g^2 (I(z_2^*)-\zeta(2))]~, \label{right helical if}
\ee
 where $ z_2=q_3^+r_2^+/(k_2^+r_1^+)$. The corresponding results for  negative helicities are obtained by complex conjugation of Eqs.~\eqref{left helical if}
and \eqref{right helical if}.

In the conformal representation,  the energy evolution parameter in Eq.~\eqref{s s contribution} is  $s_2\qs_2/(|\q_1||\q_3||\vk_1||\vk_2|)$ (instead of $s_2/(|\vk_1||\vk_2|)$, and in the two-dimensional transverse momentum space the kernel takes the form \eqref{K conformal}.
It has the representation
\be
\langle \tilde{{\cal G}_1{\cal G}_2}| {\hat K_c}|\tilde{{\cal G}'_1{\cal G}'_2}\rangle = \sum_{n=-\infty}^{+\infty}\int_{-\infty}^{+\infty}d\nu\, \omega(\nu, n)\,\langle \tilde{{\cal G}_1{\cal G}_2}|\nu, n\rangle  \langle \nu, n|\tilde{{\cal G}'_1{\cal G}'_2}\rangle~, \label{K in terms of eigenfunctions}
\ee
with the eigenfunctions  \cite{Fadin:2013hpa}
\be
\langle \tilde{{\cal G}_1{\cal G}_2}|\nu, n\rangle  =\delta(\x_1+\x_2-\q_2)\frac{1}{\sqrt{2\pi^2}}\left(\frac{r^+_1}{r_2^+}
\right)^{\frac{n}{2}+i\nu}\left(\frac{r^-_1}{r^-_2}\right)
^{-\frac{n}{2}+i\nu}\,,  \label{conformal eigenfunctions}
\ee
which form an orthonormal set with the integration measure \eqref{measure},
the eigenvalues being
\cite{Fadin:2011we}
\[
\omega (\nu , n)=\frac{g^2N_c}{8\pi^2}
\left(\frac{1}{2}\,\frac{|n|}{\nu ^2+\frac{n^2}{4}}-
\psi (1+i\nu -\frac{|n|}{2})+\psi (1-i\nu +\frac{|n|}{2}
+2\psi (1) \right)
\]
\[
\times\ \left(1-\frac{g^2N_c}{8\pi^2}\zeta (2)\right)
+\left(\frac{g^2N_c}{8\pi^2}\right)^2
\]
\[
\times\ \Biggl(\frac{1}{4}\left(\psi ^{\prime \prime}(1+i\nu +\frac{|n|}{2})+
\psi ^{\prime \prime}(1-i\nu +\frac{|n|}{2}) +
\frac{2i\nu \left(\psi '(1-i\nu +\frac{|n|}{2})-\psi '(1+i\nu
+\frac{|n|}{2})\right)}{\nu ^2+\frac{n^2}{4}}
\right)
\]
\begin{equation}
+3\zeta (3)+\frac{1}{4}\,\frac{|n|\,\left(\nu
^2-\frac{n^2}{4}\right)}{\left(\nu
^2+\frac{n^2}{4}\right)^3}\Biggr)~.
\end{equation}
Here $\psi (x)=(\ln \Gamma (x))'$.
Note that $\omega (\nu, n)$ has the important property
\begin{equation}
\omega (0,0)=0 \;,
\end{equation}
in accordance with the bootstrap conditions.
Using the representation \eqref{K in terms of eigenfunctions} and Eqs. \eqref{left helical if}, \eqref{right helical if}, we obtain  for   positive  helicities  of both  gluons
\[
\frac{t_2 \langle \tilde{G_1 R_1}|_s \;\left(e^{\hat {\cal K}_m \ln\left(\frac{s_2}{|\vk_1||\vk_2|}\right)}-1\right)|\tilde{G_2R_3}\rangle_s}{\tilde{\gamma}_{R_1R_2}\tilde{\gamma}_{R_2R_3}}
=\delta(\q_1-\vk_1-\vk_2-\q_3)g^2
(1-2\bar g^2\zeta(2))
\]
\[
\times \frac12\sum_{n=-\infty}^{+\infty}\int_{-\infty}^{+\infty}d\nu \,  \left(e^{\omega(\nu, n)\ln\left(\frac{s_2\qs_2}{|\q_1||\q_3||\vk_1||\vk_2|}\right)}-1\right) w^{\frac{n}{2}+i\nu}(w^*)
^{-\frac{n}{2}+i\nu}
\]
\[
\times\int \frac{d^2z_1}{\pi|z_1|^2}\frac{1}{1-z_1} \left(1+\bar g^2 I(z_1)\right)
z_1^{\frac{n}{2}+i\nu}(z_1^*)
^{-\frac{n}{2}+i\nu}
\]
\be
\times
\int \frac{d^2z_2}{\pi|z_2|^2}\frac{1}{1-z_2^*} \left(1+\bar g^2 I^*(z_2)\right)
(z^*_2)^{\frac{n}{2}-i\nu}z_2
^{-\frac{n}{2}-i\nu}
\ee
where $w=k_2^+q_1^+/(k_1^+q_3^+)$.

The second term in Eq.~\eqref{s s contribution} must be calculated at $D=4+2\epsilon$. Using Eqs. \eqref{left if}-\eqref{left C} and  \eqref{e L pm}  for $\langle \tilde{G_1 R_1}|_s $,
Eqs. \eqref{R IF}-\eqref{vec C R} and  \eqref{polarization} for  $|\tilde{G_2R_3}\rangle_s$,   and the results obtained in Appendix C,
we have for positive gluon helicities
\be
\frac{t_2 \langle \tilde{G_1 R_1}|_s|\tilde{G_2R_3}\rangle_s}{\tilde{\gamma}_{R_1R_2}\tilde{\gamma}_{R_2R_3}}
=\delta(\q_1-\vk_1-\vk_2-\q_3)g^2\Biggl( \frac{1}{\epsilon}+\ln\left(\frac{\vks_1
\vks_2}{\vks}\right)\Biggr)
(1-2\bar g^2\zeta(2))~.
\ee

Calculation of the second term in the right side of Eq.~\eqref{decomposition of dis s 2} is simplified  because the infrared divergency is absent in this term, since   $\langle{\cal G}_1{\cal G}_2| G_2 R_3\rangle_u$   at  $\x_2=0$ according to Eq.~\eqref{IFs - IFu}.   Therefore, we have for   positive  helicities  of both  gluons
\[
\frac{t_2 \langle \tilde{G_1 R_1}|_s \;\left(e^{\hat {\cal K}_m \ln\left(\frac{s_2}{|\vk_1||\vk_2|}\right)}-1\right)|\tilde{G_2R_3}\rangle_s}{\tilde{\gamma}_{R_1R_2}\tilde{\gamma}_{R_2R_3}}
=\delta(\q_1-\vk_1-\vk_2-\q_3)g^2
(1-2\bar g^2\zeta(2))
\]
\[
\times \frac12\sum_{n=-\infty}^{+\infty}\int_{-\infty}^{+\infty}d\nu \,  \left(e^{\omega(\nu, n)\ln\left(\frac{s_2\qs_2}{|\q_1||\q_3||\vk_1||\vk_2|}\right)}-1\right) w^{\frac{n}{2}+i\nu}(w^*)
^{-\frac{n}{2}+i\nu}
\]
\[
\times\int \frac{d^2z_1}{\pi|z_1|^2}\frac{1}{1-z_1} \left(1+\bar g^2 I(z_1)\right)
z_1^{\frac{n}{2}+i\nu}(z_1^*)
^{-\frac{n}{2}+i\nu}
\]
\be
\times\int \frac{d^2z_2}{\pi|z_2|^2}\frac{1}{1-z_2^*} \left(1+\bar g^2 I^*(z_2)\right)
(z^*_2)^{\frac{n}{2}-i\nu}z_2
^{-\frac{n}{2}-i\nu}
\ee
where $w=k_2^+q_1^+/(k_1^+q_3^+)$.

\subsection{Discontinuities in the $s_{02}, s_{13}$ and $s$ channels}
The discontinuities in the $s_{02}, s_{13}$ and $s$ channels can be expressed through the ones calculated above  with the help of the bootstrap relations \eqref{bootstrap relations}. There are four such relations,
but only three of them are independent. In general, for ${\cal A}_{2\rightarrow 2+n}$,
there are $n+2$  bootstrap relations \eqref{bootstrap relations},  for $j=0, 1, ... n+1$, but their sum is identically zero. Denoting $disc_{s_{ij}}{\cal A}^{\{-\}}_{2\rightarrow 4}/\Re {\cal A}_{2\rightarrow 4} =R^{\{-\}}_{ij}$, we have for $j=0, 1, 2$ in the relations \eqref{bootstrap relations}
\[
R^{\{-\}}_{01} +R^{\{-\}}_{02} +R^{\{-\}}_{03} =-i\pi \omega(t_1)~, \; \;
R^{\{-\}}_{12} +R^{\{-\}}_{13} -R^{\{-\}}_{01} =-i\pi\left(\omega(t_2)-\omega(t_1)\right)~, \; \;
\]
\be
 R^{\{-\}}_{23} -R^{\{-\}}_{12} -R^{\{-\}}_{02} =-i\pi\left(\omega(t_3)-\omega(t_2)\right)~. \; \;
\label{2-4 relationships}
\ee
This result gives
\[
disc_{s_{02}}{\cal A}^{\{-\}}_{2\rightarrow 4}=
disc_{s_{3}}{\cal A}^{\{-\}}_{2\rightarrow 4}-disc_{s_{2}}{\cal A}^{\{-\}}_{2\rightarrow 4} -i\pi\left(\omega(t_2)-\omega(t_3)\right){\Re {\cal A}_{2\rightarrow 4}}~, \;\;
\]
\[
disc_{s_{13}}{\cal A}^{\{-\}}_{2\rightarrow 4}=
disc_{s_{1}}{\cal A}^{\{-\}}_{2\rightarrow 4}-disc_{s_{2}}{\cal A}^{\{-\}}_{2\rightarrow 4} -i\pi\left(\omega(t_2)-\omega(t_1)\right){\Re {\cal A}_{2\rightarrow 4}}~, \;\;
\]
\[
disc_{s}{\cal A}^{\{-\}}_{2\rightarrow 4}=disc_{s_{2}}{\cal A}^{\{-\}}_{2\rightarrow 4}-
disc_{s_{1}}{\cal A}^{\{-\}}_{2\rightarrow 4}-disc_{s_{3}}{\cal A}^{\{-\}}_{2\rightarrow 4}
\]
\be
-i\pi\left(\omega(t_3)+\omega(t_1)-\omega(t_2)\right){\Re {\cal A}_{2\rightarrow 4}}~. \label{disc 02 13 s}
\ee
The same relations can be obtained from the representation of the discontinuities in terms of matrix elements of the evolution operators and the gluon production operators
between the impact factor states and  use of the bootstrap conditions \eqref{if bootstrap} -- \eqref{normalisation bootstrap} and \eqref{G bootstrap}.

Thus, all the discontinuities are expressed through  the discontinuities in $s_{1}, s_{3}$ and $s_2$ channels, and   the last one evidently  disagree with the BDS ansatz.
It is necessary to note that in the total imaginary part of ${\cal A}^{\{-\}}_{2\rightarrow 4}$ in the channel where all $s_{ij}$ are positive,  which is defined  by the sum of all the discontinuities, the contribution of the  $s_2$-channel discontinuity cancel, so we get
\be
\sum_{i=0}^{n}\sum_{j=i+1}^{n+1} disc_{s_{ij}}{\cal A}^{\{-\}}_{2\rightarrow 4}=
disc_{s_{1}}{\cal A}^{\{-\}}_{2\rightarrow 4}+disc_{s_{3}}{\cal A}^{\{-\}}_{2\rightarrow 4}-i\pi\omega(t_2)\Re {\cal A}_{2\rightarrow 4}~.
\ee

\section{Discontinuities of amplitudes with larger number of particles}

In general, there are $(n+1)(n+2)/2$  $\;s_{ij}$-channel discontinuities for the amplitude ${\cal A}_{2\rightarrow 2+n}$, that means ten discontinuities for ${\cal A}_{2\rightarrow 5}$. The bootstrap
relations \eqref{bootstrap relations} give $n+1$ connections between them.
For  ${\cal A}_{2\rightarrow 5}$ one can choose as independent   discontinuities in the channels  $s_1, \,s_2, \,s_3, \,s_4, \,$  $s_{13}$ and,  for example, $s_{04}$.
The ratios $disc_{s_{ij}}{\cal A}^{\{-\}}_{2\rightarrow 5}/\Re{\cal A}_{2\rightarrow 5}$  for the first four channels can be obtained from the results for ${\cal A}^{\{-\}}_{2\rightarrow 4}$ by evident substitutions. But the $disc_{s_{13}}{\cal A}^{\{-\}}_{2\rightarrow 5}$ contains the new matrix element
\[
\langle G_1R_1|e^{\hat {\cal K}_m \ln\left(\frac{s_2}{|\vk_1||\vk_2|}\right)} \hat{\cal G}(k_2)e^{\hat {\cal K}_m \ln\left(\frac{s_3}{|\vk_2||\vk_3|}\right)} |G_3R_4\rangle~.  \label{s 13 2-5 discontinuity}
\]
where $\hat{\cal G}(\vk_2)$ is the gluon production operator and $k_2$ is the gluon  momentum.  To calculate it one needs to know its matrix elements  $\langle{\cal G}'_1{\cal G}'_2|{\hat{\cal G}}(k_2)|{\cal G}_1{\cal G}_2\rangle$. They are known in the LO, but in the NLO only matrix elements  $\langle R_{\omega}(\q_2)|{\hat{\cal G}}(k_2)|{\cal G}_1{\cal G}_2\rangle$ are known in the ``bootstrap scheme" (see, for instance, Refs.~\cite{Kozlov:2011zza}, \cite{Kozlov:2012zza} and \cite{Kozlov:2014gaa}), which was introduced to simplify the proof of validity of the bootstrap conditions. Of course, the matrix elements $\langle{\cal G}'_1{\cal G}'_2|{\hat{\cal G}}(k_2)|{\cal G}_1{\cal G}_2\rangle$  are necessary for calculation of discontinuities of amplitudes with larger number of particles. We intend to discuss this matrix element in subsequent paper. 	

A few words about the total imaginary part of ${\cal A}^{\{-\}}_{2\rightarrow 5}$ in the channel where all $s_{ij}$ are positive. With account  of the bootstrap conditions it can be greatly simplified, so that its ratio to the real part  is expressed  through gluon trajectories and the ratios of the type shown in
Eqs.~\eqref{s_1 A24 final},  \eqref{s_3 A24 final}.

\section{Conclusion}
In this paper, using the BFKL approach, we have performed  an analysis of the discontinuities of multiple production amplitudes in invariant masses of pairs of produced gluons  in the multi-Regge kinematics. We have discovered, in particularly,  that  the discontinuities  of the four gluon production amplitudes  contradict the BDS ansatz for MHV amplitudes  in planar  ${\cal N}=4$ supersymmetric Yang-Mills theory.   This contradiction is almost obvious and is  already apparent in the leading logarithmic approximation. It appears  also in  amplitudes with more than four produced gluons.

We have obtained explicit expressions   of  all discontinuities for production of  three and four gluons, as well as of some of  discontinuities for  production of  a greater number of gluons in the next-to-leading logarithmic  approximation.  It turns out that certain  discontinuities have a rather complicated form. In particular,  their   colour structure
differs from the colour structure of the real part of the corresponding  amplitude.  In the sum of all discontinuities the complicated pieces cancel due to the bootstrap conditions, so that the sum acquires a relatively simple form  and  the same colour structure as the real part. This result can be important for further development of the BFKL  approach.

\newpage
\setcounter{equation}{0}
\def\theequation{A.\arabic{equation}}
\section*{Appendix A}

First, consider  Eq.~\eqref{rel 1}.
Using Eq.~\eqref{R omega} and Eqs.~\eqref{K representation}, \eqref{K = K r + omega} and \eqref{K B r} we have
\[
\langle R_{\omega}|^B\left[ \ln\left({\hat{\vec r}_1^{\,2}}{\hat{\vec r}_2^{\,2}}\right),{\hat{\cal K}}_r^B\right]|{\cal G}_1{\cal G}_2\rangle
=\frac{\bar g^2\delta(\x_1+\x_2-\q)}{\Gamma(1-\epsilon)\pi^{1+\epsilon}} T^R_{{\cal G}_1{\cal G}_2}\int\frac{d\xp_1d\xp_2\delta(\xp_1+\xp_2-\q)}{\xps_1\xps_2}
\]
\be
\times\left(\frac{\xs_1\xps_2+\xs_2\xps_1}{(\x_1-\xp_1)^2} -\qs\right)\ln\left(\frac{\xps_1\xps_2}{\xs_1\xs_2}\right)~. \label{int 1}
\ee
Due to the symmetry under the $\x_1\leftrightarrow \x_2, \;\;\xp_1\leftrightarrow \xp_2$ exchange, it is sufficient to calculate  in Eq.~\eqref{int 1} the integral with
$\ln\left(\frac{\xps_1}{\xs_1}\right)$ an to add in the answer  the term with  $\x_1\leftrightarrow \x_2$.
The integral is not infrared divergent and can be evaluated at $\epsilon =0$. It can be done using the decomposition
\[
\frac{1}{\xps_1\xps_2}\left(\frac{\xs_1\xps_2+\xs_2\xps_1}{(\x_1-\xp_1)^2} -\qs\right)
=\left(\frac{1}{r_1^{'+}}+\frac{1}{r_1^{+}-r_1^{'+}}\right)\left(\frac{1}{r_1^{-}-r_1^{'-}}-\frac{1}{q^--r_1^{'-}}\right)
\]
\be
 +\left(\frac{1}{r_1^{'-}}+\frac{1}{r_1^{-}-r_1^{'-}}\right)\left(\frac{1}{r_1^{+}-r_1^{'+}}-\frac{1}{q^+ -r_1^{'+}}\right)~, \label{decomposition}
\ee
and the integral
\[
\int\frac{d \vec l}{\pi}\left(\frac{1}{( a^+ -  1^+)}
\frac{1}{( b^- -  1^-)}+\frac{1}{( a^- -  1^-)}
\frac{1}{( b^+-  1^+)}\right)\ln\left(\frac{\vec l^{\;2}}{\mu^2}\right)
\theta(\Lambda^2-\vec l^{\;2})
\]
\begin{equation}
=\ln\left(\frac{\Lambda^2}{(\vec a -\vec b)^{2}}\right)
\ln\left(\frac{\Lambda^2(\vec a -\vec b)^{2}}{\mu^4}\right) +\ln\left(\frac{(\vec a -\vec b)^{2}}{\vec b^{\;2}}\right)
\ln\left(\frac{(\vec a -\vec b)^{2}}{\vec a^{\;2}}\right)~. \label{int master}
\end{equation}
The upper integration limit $\Lambda$ is introduced because the separate terms of the decomposition  \eqref{decomposition} give divergent integrals. In the sum the divergencies cancel and that leads to the result
\[
\int\frac{d\xp_1d\xp_2\delta(\xp_1+\xp_2-\q)}{\xps_1\xps_2}
\left(\frac{\xs_1\xps_2+\xs_2\xps_1}{(\x_1-\xp_1)^2} -\qs\right)\ln\left(\frac{\xps_1\xps_2}{\xs_1\xs_2}\right)
\]
\be
=  -2\pi \ln\left(\frac{\xs_1}{\qs}\right)\ln\left(\frac{\xs_2}{\qs}\right)~.
\ee
Using this result in Eq.~\eqref{int 1} we come to Eq.~\eqref{rel 1}.

\setcounter{equation}{0}
\def\theequation{B.\arabic{equation}}
\section*{Appendix B}
Let us consider the integral in Eq.~\eqref{int C}. The piece of this integral with the term $\q_2$
is known from the calculation of
$\omega^B(t_1)$ and gives (with $\vk=\q_1-\q_2$)
\be
\q_2\int d\x_1d\x_2\frac{\qs_1 }{\xs_1\xs_2}\delta(\x_1+\x_2 -\q_1)  = 2\pi^{1+\epsilon}\Gamma(1-\epsilon){\q_2}\left(
\frac{1}{\epsilon} +\ln\qs_1\right)~. \label{int q 2}
\ee
The integral with the second term
can be represented as
\[
\int d\x_1d\x_2\frac{\qs_1 }{\xs_1\xs_2}\delta(\x_1+\x_2 -\q_1)  (\x_1-\q_2)\frac{\qs_2}{(\x_1-\q_2)^2}
\]
\be
=\frac{\qs_1 \qs_2}{2}\frac{\partial}{\partial \q_2}\int \frac{d\x_1}{\xs_1(\x_1-\q_1)^2} \ln(\x_1-\q_2)^2~.  \label{int x 1 - q 2}
\ee
The  last integral can be  written  as sum of two integrals:
\be
\int \frac{d\x_1}{\xs_1(\x_1-\q_1)^2} \ln(\x_1-\q_2)^2 =
\frac12\int \frac{d\vl}{(\q_2-\vl)^2(\vk +\vl)^2} \left(\ln\left(\frac{\vls}{\qs_2}\frac{\vls}{\vks}\right)+\ln\left(\qs_2\vks\right)\right)~.
\ee
Here the second integral is known, whereas in the first one the contributions of the singularities at $(\q_2-\vl)=0$ and $(\vk+\vl)=0$ cancel and the integral can be calculated at $\epsilon =0$ using the decomposition
\be
\frac{1}{(\q_2-\vl)^2(\vk +\vl)^2}= \frac{1}{\qs_1}\left(\frac{1}{q_2^+-l^+}+\frac{1}{k^+ +l^+}\right)\left(\frac{1}{q_2^--l^-}+\frac{1}{k^- +l^-}\right)
\ee
and the integral \eqref{int master}. As a result, we have
\[
\int \frac{d\x_1}{\xs_1(\x_1-\q_1)^2} \ln(\x_1-\q_2)^2 = \pi^{1+\epsilon}\Gamma(1-\epsilon)\frac{1}{\qs_1}\left(\ln\left({\qs_2\vks}\right)
\left(\frac{1}{\epsilon} +\ln \qs_1 \right) +\frac12 \ln^2\left(\frac{\vks}{\qs_2}\right)\right)~. \label{int log}
\]
Substituting  this result in Eq.~\eqref{int x 1 - q 2}  and using  Eq.~\eqref{int q 2}, we obtain
\[
\int d\x_1d\x_2\frac{\qs_1 }{\xs_1\xs_2}\delta(\x_1+\x_2 -\q_1)\left(\q_2
+(\x_1-\q_2)\frac{\qs_2}{(\x_1-\q_2)^2}\right)
\]
\be
={\pi^{1+\epsilon}\Gamma(1-\epsilon)}\left(\q_2
 +\vk\frac{\qs_2}{\vks}\right)\left(
\frac{1}{\epsilon} +\ln\left(\frac{\qs_1\vks}{\qs_2}\right)\right)~. \label{int C A}
\ee

\setcounter{equation}{0}
\def\theequation{C.\arabic{equation}}
\section*{Appendix C}
Let us calculate now the chiral components of the tensor
\be
J^{ij}=\frac{1}{{\pi^{1+\epsilon}\Gamma(1-\epsilon)}}\int   \frac{d\x}{\xs(\q_2-\x)^2}\Bigl(\frac{q_1}{\qs_1}-\frac{(\q_1-r)}{(q_1-\x)^2}\Bigr)^i\Bigl(\frac{q_3}{\qs_3}-\frac{(\q_3-r)}{(q_3-\x)^2}\Bigr)^j \label{J ij}~.
\ee
Writing
\be
\Bigl(\frac{q_1}{\qs_1}-\frac{(\q_1-r)}{(q_1-\x)^2}\Bigr)^i= \Bigl(\frac{q_1}{\qs_1}-\frac{k_1}{\vks_1}\Bigr)^i + \Bigl(\frac{k_1}{\vks_1}-\frac{(\q_1-r)}{(q_1-\x)^2}\Bigr)^i
\ee
we can split the tensor in the sum of two pieces:
\be
J^{ij}=J_1^{ij}+J_2^{ij},
\ee
where
\[
J_1^{ij}=\Bigl(\frac{q_1}{\qs_1}-\frac{k_1}{\vks_1}\Bigr)^i\frac{1}{{\pi^{1+\epsilon}\Gamma(1-\epsilon)}}\int  \frac{d\x}{\xs(\q_2-\x)^2}
\Bigl(\frac{q_3}{\qs_3}-\frac{(\q_3-r)}{(q_3-\x)^2}\Bigr)^j ,
\]
\be
J_2^{ij}=\frac{1}{{\pi^{1+\epsilon}\Gamma(1-\epsilon)}}\int \frac{d\x}{\xs(\q_2-\x)^2}
\Bigl(\frac{k_1}{\vks_1}-\frac{(\q_1-r)}{(q_1-\x)^2}\Bigr)^i
\Bigl(\frac{q_3}{\qs_3}-\frac{(\q_3-r)}{(q_3-\x)^2}\Bigr)^j .
\ee
The first  tensor can be obtained from Eq.~\eqref{int C A} by the replacement $\q_1\rightarrow \q_2, \; \q_2\rightarrow \q_3$; we get
\be
J_1^{ij} \simeq \Bigl(\frac{q_1}{\qs_1}-\frac{k_1}{\vks_1}\Bigr)^i\Bigl(\frac{q_3}{\qs_3}+\frac{k_2}{\vks_2}\Bigr)^j\frac{1}{\qs_2}\Biggl( \frac{1}{\epsilon}+\ln\left(\frac{\qs_2
\vks_2}{\qs_3}\right)\Biggr).
\ee
The tensor $J_2^{ij}$ is infrared finite and can be calculated at $\epsilon =0 $. The calculation of its   chiral components can be  performed easily using the decomposition of the integrand   into a  sum of terms of the type  $(a^+-r^+)^{-1}(b^--r^-)^{-1}$ and the integral
\be
\int\frac{d^2r}{\pi(a^+-r^+)(b^--r^-)}\theta(\Lambda^2-\xs) =\ln\left(\frac{\Lambda^2}{(\vec a-\vec b)^2}\right)~.
\ee
It gives
\[
J_2^{++} = \Bigl(\frac{k}{\vks}-\frac{k_1}{\vks_1}\Bigr)^+\Bigl(\frac{q_3}{\qs_3}+\frac{k_2}{\vks_2}\Bigr)^+\frac{1}{\qs_2}\ln\left(\frac{\qs_3\vks_1}{\vks_2\qs_1}\right),\]
\be
J_2^{+-} = \Bigl(\frac{q_1}{\qs_1}-\frac{k_1}{\vks_1}\Bigr)^+\Bigl(\frac{q_3}{\qs_3}+\frac{k_2}{\vks_2}\Bigr)^-\frac{1}{\qs_2}\ln\left(\frac{\qs_3\vks_1}{\vks\qs_2}\right),
\ee
where $\vk=\vk_1+\vk_2$ and $J_2^{--}= \left(J_2^{++}\right)^*, \;\;J_2^{-+}= \left(J_2^{+-}\right)^*.$ Therefore, for the $+-$ component we have
\[
J^{+-} \simeq \Bigl(\frac{q_1}{\qs_1}-\frac{k_1}{\vks_1}\Bigr)^+\Bigl(\frac{q_3}{\qs_3}+\frac{k_2}{\vks_2}\Bigr)^-\frac{1}{\qs_2}\Biggl( \frac{1}{\epsilon}+\ln\left(\frac{\vks_1
\vks_2}{\vks}\right)\Biggr)
\]
\be
=-\frac{1}{q_1^-k_1^-q_3^+k_2^+}\Biggl( \frac{1}{\epsilon}+\ln\left(\frac{\vks_1
\vks_2}{\mu^2\vks}\right)\Biggr). \label{J +-}
\ee


\end{document}